\documentclass[fleqn,usenatbib,usedcolumn]{mnras}
\usepackage[utf8]{inputenc}
\usepackage{graphicx}	
\usepackage{amsmath}
\usepackage{amssymb} 
\usepackage{multicol}
\usepackage{bm}
\usepackage{verbatim}
\usepackage{wrapfig}
\usepackage{multirow}

\usepackage{pdflscape}
\usepackage[]{todonotes}
\usepackage[]{algorithm2e}

\newcommand{\OHb}{$\log$ [OIII]/H$\beta$}
\newcommand{\NHa}{$\log$ [NII]/H$\alpha$}
\newcommand{\EWH}{$\log$ EW(H${\alpha}$)}

\newcommand{\Ha}{H$\alpha$}
\newcommand{\Hb}{H$\beta$}
\newcommand{\Oiii}{[OIII] $\lambda$5007}
\newcommand{\Nii}{[NII] $\lambda$6583}

\newcommand{\VEC}[1] {{\boldsymbol{{ #1}}}}

\usepackage[T1]{fontenc}
\usepackage{ae,aecompl}
\usepackage{newtxtext,newtxmath}
\title[Emission-line galaxy classification with GMM]{A probabilistic approach to emission-line galaxy classification}

\author[R.~S.~de Souza et al.]
{R.~S.~de Souza$^{1,2,3}$\thanks{drsouza@ad.unc.edu}, M.~L.~L.~Dantas$^{3}$, M.~V.~Costa-Duarte$^{3,4}$,
E.~D.~Feigelson$^{5}$,
\newauthor  M.~Killedar$^{6}$, P.-Y.~Lablanche$^{7,8}$,  R.~Vilalta$^{9}$, A.~Krone-Martins$^{10}$,   R.~Beck$^{11}$, F. Gieseke$^{12}$, 
\newauthor for the COIN collaboration\\
$^{1}$Department of Physics \& Astronomy, University of North Carolina at Chapel Hill, Chapel Hill, NC 27599-3255, USA\\
$^{2}$MTA E\"otv\"os University, EIRSA ``Lendulet'' Astrophysics Research Group, Budapest 1117, Hungary\\
$^{3}$Instituto de Astronomia, Geof\'isica e Ci\^encias Atmosf\'ericas, Universidade de S\~ao Paulo, 
R. do Mat\~ao 1226, 05508-090, S\~ao Paulo, Brazil\\
$^{4}$Leiden Observatory, Leiden University, Niels Bohrweg 2, 2333 CA Leiden, The Netherlands. \\
$^{5}$Dept. of Astronomy \& Astrostatistics and Center for Astrostatistics, Penn State University, 525 Davey Laboratory, University Park PA 16802, USA \\
$^{6}$Burnet Institute, 85 Commercial Road, Melbourne, VIC 3004, Australia\\
$^{7}$Stellenbosch University, Matieland 7602, South Africa\\
$^{8}$African Institute for Mathematical Sciences, 6 Melrose Road, Muizenberg, Cape Town 7945, South Africa \\
$^{9}$Department of Computer Science, University of Houston
4800 Calhoun Rd, Houston TX 77204-3010, USA\\
$^{10}$CENTRA/SIM, Faculdade de Ci\^encias, Universidade de Lisboa, Ed. C8, Campo Grande, 1749-016, Lisboa, Portugal\\
$^{11}$Department of Physics of Complex Systems, E\"otv\"os Lor\'and University, Budapest 1117, Hungary \\
$^{12}$Radboud University Nijmegen, Toernooiveld 212, The Netherlands 
}
\date{Last updated ; in original form }

\pubyear{2017}

\begin{document}
\maketitle

\begin{abstract}
We invoke a Gaussian mixture model (GMM) to  jointly analyse two traditional emission-line classification schemes of galaxy ionization sources: the Baldwin-Phillips-Terlevich (BPT) and $\rm W_{H\alpha}$ \emph{vs.} [NII]/H$\alpha$ (WHAN) diagrams, using spectroscopic data from the Sloan Digital Sky Survey Data Release 7 and SEAGal/STARLIGHT datasets.  
We apply a GMM to empirically define classes of galaxies in a three-dimensional space spanned by the \OHb, \NHa, and \EWH \, optical parameters. The best-fit GMM based on several statistical criteria suggests a solution around four Gaussian components (GCs), which are capable to explain up to 97 per cent of the data variance. Using elements of information theory, we compare each GC to their respective astronomical counterpart. GC1 and GC4 are associated with star-forming galaxies, suggesting the need to define a new starburst subgroup. GC2 is associated with BPT's Active Galaxy Nuclei (AGN) class and WHAN's weak AGN class. GC3 is associated with BPT's composite class and WHAN's strong AGN class. Conversely, there is no statistical evidence -- based on four GCs -- for the existence of a Seyfert/LINER dichotomy in our sample. Notwithstanding, the inclusion of an additional GC5 unravels it.  The GC5 appears  associated to the LINER and Passive galaxies on the BPT and WHAN diagrams respectively. This indicates that if the  Seyfert/LINER dichotomy is there, it does not account significantly to the global data variance and may be overlooked by standard metrics of goodness of fit. 
Subtleties aside, we demonstrate the potential of our methodology to recover/unravel different objects inside the wilderness of astronomical datasets, without lacking the ability to convey physically interpretable results. The probabilistic classifications from the GMM analysis are publicly available within the COINtoolbox \href{https://cointoolbox.github.io/GMM\_Catalogue/}{https://cointoolbox.github.io/GMM\_Catalogue/}.
\end{abstract}



\section{Introduction}

Classification of objects has long been recognized as a major driver in 
natural sciences, from taxonomical classification of species, 
anthropological variation of  cultures \citep[e.g.][]{stocking1968race}, 
to the vastness of galaxy shapes \citep{DeVaucouleurs1959}. Empirical 
classifications are powerful triggers for novel theories, an archetypal 
example being the Linnaean classification of organisms 
\citep{linnaeus1758} that subsequently inspired the birth of 
Darwin’s renowned theory of common descent  \citep{darwin1859origin}. 

Even though the properties of objects in nature may lie along a 
continuum, and groups may be defined by fuzzy boundaries, it may still be 
practical to divide them into categories that ideally reflect some 
physical distinctions. In astronomy, a canonical example is the one-dimensional Morgan-Keenan \citep{MK1973} system  of spectral  stellar 
classification, in which stars of each class share similar ionization 
states or effective temperatures. The system was later used to compose 
the two-dimensional Hertzsprung-Russell  diagram \citep[e.g.][and references therein]{Chiosi1992}, in which  different 
stages of stellar evolution (e.g. main sequence, white dwarfs, giants, 
etc.) are grouped according to their luminosity (or magnitude) and 
effective temperature (or colour).

In the context of extragalactic astrophysics, various classification schemes have been proposed to help ascertain the main drivers regulating galaxy evolution; this task becomes imperative in the face of the deluge of information gathered by current \citep[e.g. Sloan Digital Sky Survey,][]{York2000,Zhang2015} and upcoming \citep[e.g. Large Synoptic Survey Telescope,][]{ivezic2008lsst} large-scale sky surveys. Some examples are the  classification of galaxies based on their morphological type \citep{Lintott2008}, their surrounding environment \citep{vonderLinden2010}, or their spectral features \citep{Morgan1957,Ucci2017}.

Notably, the collisionally excited emission--lines are powerful diagnostics to differentiate  galaxies according to their ionization power source \citep[e.g.][]{Stasinska2007}, i.e. nuclear emission, star formation, and so forth. Some of the most widely used emission-line diagnostics are the Baldwin-Phillips-Terlevich \citep[BPT;][]{baldwin81,VeilleuxOsterbrock1987,rola1997,kewley2001,kauffmann2003,Stasinska2006,Schawinski2007} and, more recently, the $\rm W_{H\alpha}$ \emph{vs.} [NII]/H$\alpha$  \citep[WHAN;][]{Cid2010,CidFernandes2011} diagrams.

The BPT diagram\footnote{Also known as the \emph{Seagull} diagram.} classifies galaxies into star-forming (SF), composite, and active galactic nuclei (AGN) hosts. The latter can be further subdivided into low-ionization nuclear emission-line region (LINER) galaxies and Seyferts. The lines used to define such classification are \Hb, \Oiii, \Ha, and \Nii, and the galaxies are classified in the parameter space formed by \NHa~($x$-axis) and \OHb~($y$-axis). SF galaxies are those in which the photoionization processes responsible for the emission-lines are mainly due to young hot stars; they reside mostly in the left wing locus of the BPT diagram. On the opposite side lie the AGN-dominated objects, which are composed of two large groups, the LINERs and the Seyferts, usually divided by the line proposed by \cite{Schawinski2007}. Objects classified as LINERs have an uncertain source of photoionization \citep{Belfiore2016}, that could be due to true nuclear activity or perhaps evolved stellar populations \citep{Singh2013}. Finally, we have the composite area of the diagram, marking the transition between SF and AGN objects, which are usually delimited by the theoretical extreme starburst line described by \cite{kewley2001} and the empirical starburst line proposed by \cite{kauffmann2003}. It is worth noting that these boundaries are still a matter of debate and further alternative lines have been proposed \citep[for instance][]{Stasinska2006}.

The WHAN diagram has the same emission line ratio (i.e. \NHa) on the $x$-axis as the BPT. On the other hand, it uses the equivalent width of H$\alpha$, i.e. \EWH, as the characteristic parameter on the $y$-axis, instead of \OHb. The WHAN diagram uses a set of perpendicular and parallel straight lines to divide galaxies into strong AGN (sAGN), weak AGN (wAGN), SF, and retired/passive galaxies. Because H$\alpha$ is also used for the $y$-axis, a larger number of galaxies may be analysed, many of which would not appear on the BPT due to the lack of some emission features (\Hb~ and \Oiii), i.e. groups of galaxies mainly represented by the \textit{retired} and \textit{passive} galaxy classes \citep{Cid2010,CidFernandes2011,Stasinska2015}. 
However, it lacks the definition of a transitional \emph{composite} region.

Other examples of emission-line diagrams are: the Mass-Excitation diagram \citep{Juneau2014}, which also uses the stellar mass of galaxies as a proxy for classification; the Colour-Excitation diagram \citep{Yan2011}; the Blue diagram \citep{Lamareille2004,lamareille2010}; and the Trouille-Barger-Tremonti diagram \citep{Trouille2011}. Moreover, many of these  classification methods also include photometric information, such as the mid-infrared colour-colour diagrams \citep{Lacy2004,Sajina2005,Stern2005}. More recently, classifications based on UV information have been proposed in order to better understand the nature of galaxies at high redshifts \citep{Feltre2016}.

A common characteristic of most of these diagrams and the majority of standard classification systems in astronomy is the sharp division between classes, in which boundaries are more often than not defined by eye or fitted without accounting for a smooth  transition between objects. Given the  ever-increasing richness of information enclosed in astronomical surveys, we advocate updating standard classification schemes under the paradigm of contemporary statistical methods, while  still maintaining the crucial role of expert knowledge in the physical interpretation of data-driven classes.

From a methodological point of view, a recent trend in object classification has been the reliance on \emph{machine learning} for data analysis~\citep{HastieTF2009,Murphy2012}. While being conceptually very similar to well-known existing statistical methods, the tremendous increase in both available data as well as computational power over the past two decades has led to the development of a variety of advanced techniques. Machine learning aims at deriving ``models'', which can retrieve useful information in an automatic manner. Examples of machine learning in astronomy are: supernovae classification \citep[e.g.][]{Richards2012,Ishida2013,Karpenka2013,Lochner2016,Sasdelli2016MNRAS}, studies of emission-line spectra of galaxies \citep{Beck2016,Ucci2017}, photometric redshift estimation \citep[e.g.][]{Collister2004,Krone-Martins2014,Cavuoti2015,Elliott2015,Hogan2015,Beck2017}, and detection of galaxy outliers \citep[e.g.][]{Baron2017}. 

In the past few decades, various new techniques have been proposed along two prominent lines of research:  supervised and unsupervised learning~\citep{HastieTF2009}. For the former, one is given labelled data, i.e. objects described by a set of  parameters along with a class label (e.g. discrimination between early and late type galaxies based on their photometric colours).  The other line of research, unsupervised learning, 
does not make assumptions about pre-existing labels; the goal is to automatically derive conclusions about the data structure by assigning  ``similar'' objects to the same group. Thus, 
in contrast to supervised classification, no information is made available about the categories or classes to which objects belong. Instead, the unsupervised learning model must discover such classes. While supervised learning methods have been applied previously to the specific problem of AGN classification \citep{Beck2016}, the unsupervised approach is by definition more suitable for challenging or reinforcing the existing classification paradigm, as it can study what statistical evidence the measurement data contain in support of given classes. For this reason, in this paper we adopt an unsupervised approach.

One type of unsupervised model is the so-called Gaussian mixture model \citep[GMM; e.g.][]{everitt2011cluster}.  Most unsupervised clustering methods, like the popular friends-of-friends algorithm \citep[][also known as single-linkage agglomerative clustering in statistical parlance]{Davis1985}, are non-parametric and less robust against different choices of algorithms \citep{Feigelson2012msma}.  GMMs, in contrary are parametric, hence solvable by maximum likelihood approach. This makes the  GMMs a \textit{desideratum}  due to its  objective,  stable and  interpretable  probabilistic results.  

Previous examples of the application of mixture models in astronomy are the search for sub-cluster structures of young stars in massive star-forming regions \citep{Kuhn2014}, and the separation of millisecond pulsars from a broader sample \citep{Lee2012}.
This paper demonstrates the application of GMM for the emission-line classification of galaxies, and further discusses how the data-driven groups  can  be related to classic classifications, which are based on expert domain knowledge.

The outline of this  paper is as follows. In Section~\ref{sec:data} we provide an overview of the  sample selection. Section~\ref{sec:method} describes the GMM methodology.  We present our main results  in Section~\ref{sec:res}, discuss their physical meaning in Sections~\ref{sec:cluster_assessment} and \ref{sec:astrophysical_discussion}, and present our conclusions in  Section~\ref{sec:results_discussion}. 
The standard $\Lambda$CDM cosmology with \{$H_0$, $\Omega_M$, $\Omega_{\Lambda}\}$ = \{70 $\mathrm{km\,s^{-1}Mpc^{-1}}$, 0.3, 0.7\} has been used throughout the paper.

\section{Catalogue}
\label{sec:data}

The galaxy sample used in this work is the result of matching two databases: the Sloan Digital Sky Survey Data Release 7 \citep[SDSS-DR7,][]{abazajian2009} and the public SEAGal/STARLIGHT catalogue\footnote{\url{http://casjobs.starlight.ufsc.br/casjobs/}}.
The SDSS-DR7 comprises photometry in five broad-band filters ($ugriz$) and optical spectroscopy between 3800\AA $\,$ and 9200\AA $\,$ in the observed frame. 
Our initial  sample retrieved from the  SDSS-DR7 database is volume-limited\footnote{This choice follows a similar procedure as e.g. \citet{Mateus2006,CidFernandes2011}, and aims to  mitigate the  bias towards galaxies with  the presence of strong emission lines.} and composed of galaxies brighter than $M_r<-19.88+5\log h_{70}$, with $h_{70} \equiv H_0/70\,\mathrm{km\,s^{-1}Mpc^{-1}}$,  over the redshift range $0.015 < z < 0.075$\footnote{The narrow redshift range herein employed has the aim to mitigate potential biases caused by evolutionary effects.}. The  magnitudes in our sample were treated to account for the effects of Galactic extinction, and  further $K$-corrected using the software \textsc{kcorrect} v3.2 \citep{Blantonetal2003}. 

The SEAGal/STARLIGHT catalogue provides spectral synthesis parameters such as average stellar metallicity ($\langle Z/Z_{\odot} \rangle_{L}$, with respect to the Sun's metallicity), average stellar age ($ \langle \log (t/{\rm yr}) \rangle _{L}$, in units of year), and the 4000 \AA~ break (D$_n$4000)\footnote{For more information, we refer the reader to the STARLIGHT Casjobs Schema Browser: \url{http://casjobs.starlight.ufsc.br/casjobs/field_list.html}.}, as well as emission-line measurements of all SDSS-DR7 galaxies. The empirical spectral synthesis technique is carried out using the \textsc{starlight} code \citep{CidFernandesetal2005}, which fits the stellar continuum by using a library of simple stellar populations from \cite{BC03}.

\begin{figure*}
\includegraphics[clip,width=0.45\linewidth]{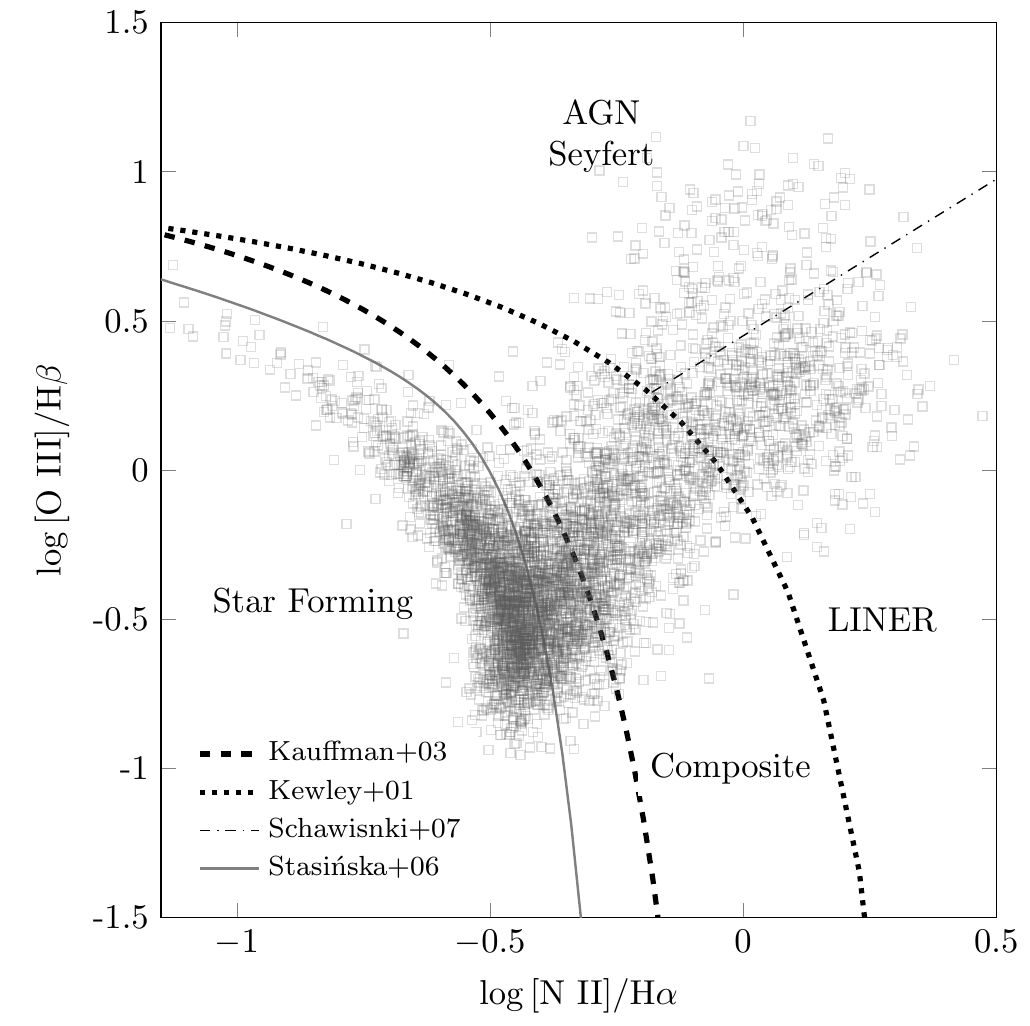}
\includegraphics[clip,width=0.45\linewidth]{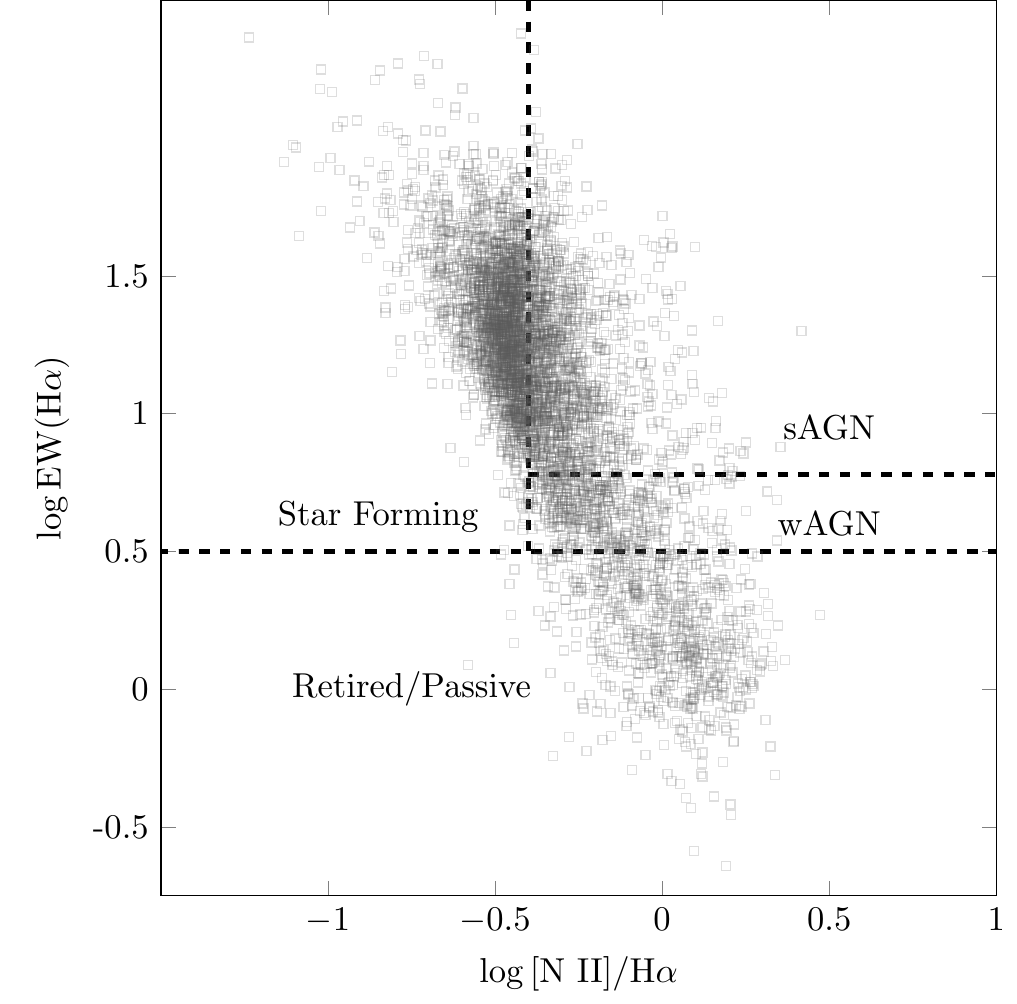}
\caption{BPT and WHAN diagrams, from left to right, with galaxy points from the SDSS and SEAGal datasets.
On the BPT diagram, the curves define the division between SF and AGN classes (dotted: \citealt{kewley2001}; solid: \citealt{Stasinska2006}; dashed: \citealt{kauffmann2003}), and the dot-dashed line shows the division between AGN and LINERs as suggested by \citet{Schawinski2007}. 
On the WHAN diagram, the dashed straight lines discriminate between sAGN, wAGN, SF, and retired/passive galaxies \citep{CidFernandes2011}.  For  better visualization,  the points in the figure represent a sub-sample of 10,000 randomly selected galaxies.}
\label{fig:bptwhan}
\end{figure*}

The emission-lines are fitted by subtracting the stellar continuum and using a Gaussian profile \citep[for more details, see][]{Cid2010}. For this analysis, fluxes and equivalent widths of the \Oiii, \Hb, \Ha\ and \Nii\ emission-lines are extracted from the SEAGal/STARLIGHT database. In order to ensure good quality measurements, we impose a signal-to-noise ratio of $S/N > 3$ and a fraction of bad pixels lower than 
$25\%$, for all emission-lines.
The aforementioned constraints and matching between both databases lead to a final galaxy sample that consists of 83,578 objects. Figure~\ref{fig:bptwhan} displays the projections of all galaxies in the sample on the traditional BPT and WHAN diagrams.

\section{Gaussian mixture models}
\label{sec:method}

GMM is a parametric model that, within a given feature space, assumes the existence of classes  that can be described by a superposition of  multivariate Gaussian distributions  \citep[e.g.][]{McLachlan00,hastie01,mengersen2011,Murphy2012}. 
The model is defined as a  probability density function 
comprised of a weighted summation of Gaussian component (GC) densities.  The goal is to describe the distribution of data in a certain feature space and assign probabilities to the membership of a given datum in each class. 
More specifically, for a total of $K$ clusters in a $d$-dimensional parameter space, the GMM is a probability distribution $p(x)$ given by a weighted summation  of $K$ components:
\begin{equation}
p(x)  = \sum_{\kappa =1}^k \zeta_k\phi(x;\VEC{\mu}_\kappa,\VEC{\Sigma}_\kappa),
\label{eq:px}
\end{equation}
with mixture weights denoted by $\zeta_\kappa$, and $\sum \zeta_\kappa = 1$. Here, each of the $K$ model components is described as a $d$-variate Gaussian density, fully characterized by its mean $\VEC{\mu}_\kappa$ and covariance matrix $\VEC{\Sigma}_\kappa$:
\begin{equation}
\phi(x;\VEC{\mu}_\kappa,\VEC{\Sigma}_\kappa) = \frac{1}{\sqrt{(2\pi)^d|\VEC{\Sigma}_\kappa|}}e^{-\frac{1}{2}(x-\VEC{\mu}_\kappa)\VEC{\Sigma}_\kappa^{-1}(x-\VEC{\mu}_\kappa)}.
\end{equation}
Various ways exist to fit such a GMM to a given set of data points, among which stands out the popular  Expectation Maximization algorithm \citep[EM;][]{Dempster77,Hoboken08}. This work adopts the EM algorithm from the \textsc{r} \citep{rlanguage}  package \textsc{mclust} \citep{mclust} to fit the GMMs.\footnote{Additionally, an independent GMM was implemented using the  \textsc{python} package \textsc{scikit-learn} \citep{scikit-learn} to check the cross-consistency of our results.}   
 
\section{GMM application to SDSS and SEAGal data}
\label{sec:res}

This section presents the results from the application of a GMM to the SDSS-DR7 and SEAGal/STARLIGHT catalogues.

\subsection{Results}

We now apply the GMM to our galaxy catalogue projected into the joint combination of the BPT and WHAN diagrams. Hence, the  dimension of the parameter space is $d = 3$ and the data vector \VEC{x} in Equation \ref{eq:px}  is given  by:
\begin{equation}
\VEC{x} = \left(\begin{array}{c} $\NHa$ \\  $\OHb$ \\
$\EWH$\end{array}\right).
\end{equation}
The output is a soft classification of each object given by the membership probability for each group, together with parameters  $\VEC{\mu}_\kappa$ and $\VEC{\Sigma}_\kappa$ for each 3-variate GC.

\begin{figure*} 
\includegraphics[width=0.95\linewidth]
{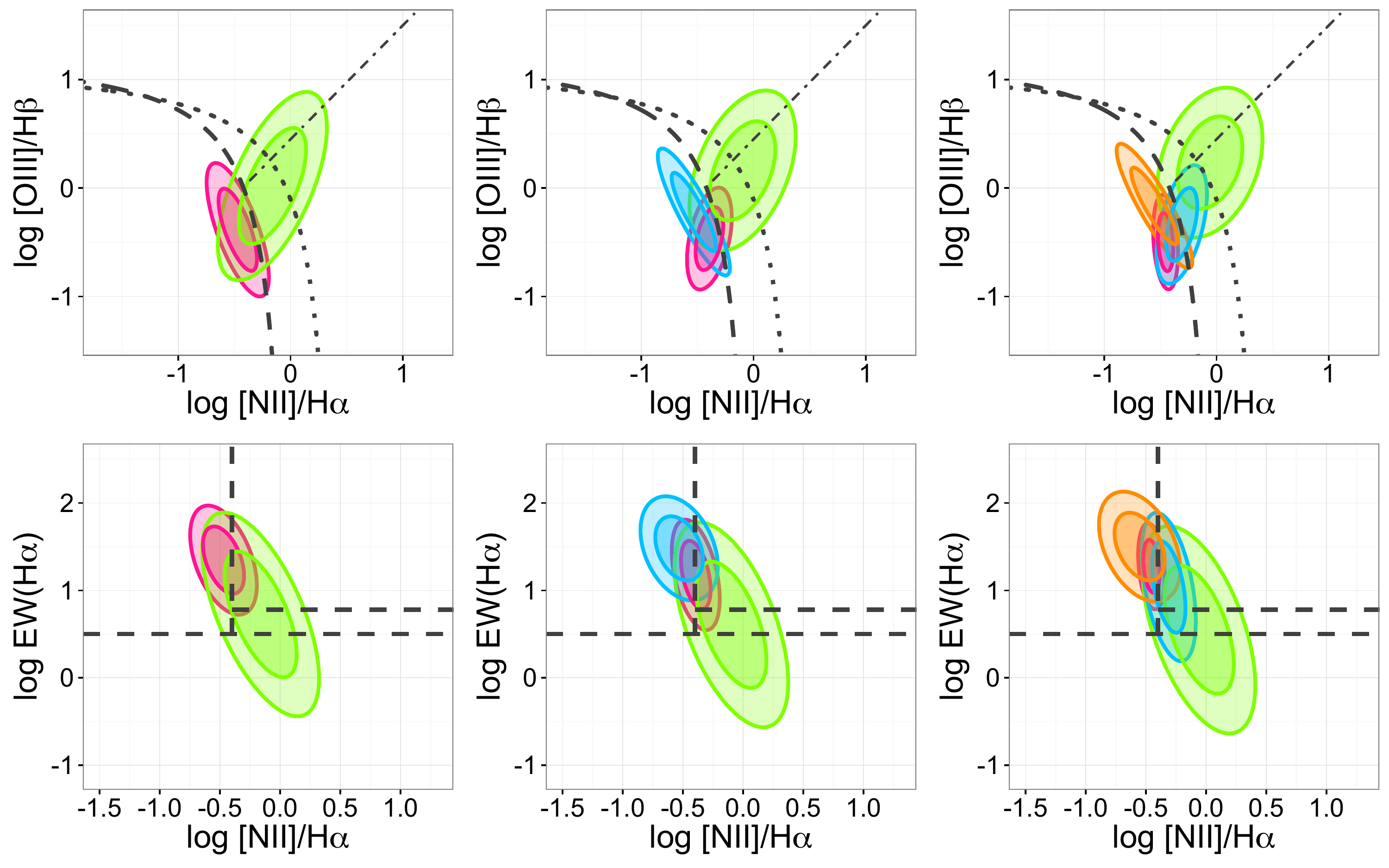}
\caption{The Gaussian components  projected onto the BPT (top panels) and WHAN (bottom panels) diagrams. From left to right are the solutions for 2, 3 and 4 GCs. For each component the thick lines represent 68\% and 95\%  confidence levels, respectively.
}
\label{res:GMM_2d_BPT_WHAN}
\end{figure*}

We show the results in Figure~\ref{res:GMM_2d_BPT_WHAN} of the two-, three- and four-cluster solutions, each projected onto the two-dimensional BPT and WHAN parameter space. A visual inspection suggests that  while the whole population of galaxies cannot be explained by a single Gaussian distribution, their overall distribution can be approximated by multiple Gaussian clusters. The contours represent  $68\%$ and $95\%$ confidence levels around the mean of each GC respectively. The leftmost panel of Figure~\ref{res:GMM_2d_BPT_WHAN} shows that the two-cluster solution roughly separates the SF and AGN dominated galaxies in both diagrams. The solution with three clusters, displayed in the middle panel, identifies a composite region of the BPT diagram, and a possible transitional  region in the  WHAN  diagram, which will be further discussed in Section \ref{sec:cluster_assessment}. The solution with four clusters indicates a possible subdivision of the SF region in the BPT, which may be connected to the existence of starburst galaxies, predominantly located in the top-left region of the BPT diagram. The parameters for the four-cluster solution are shown in Table~\ref{tab:para}.  Four GCs are preferred to describe the galaxy population in the \NHa, \OHb, and \EWH \, feature space, based on a set of cluster validation methods, as described next.

\begin{table}
\caption{Parameters of the GMM solution with four Gaussian components for the galaxy distribution in the \NHa, \OHb, and \EWH \, space. Shown are the mixture weights  $\zeta_{1}\dots \zeta_{4}$, central  vectors  of the clusters,
$\VEC{\mu}_{1} \dots \VEC{\mu}_{4}$,  and covariance matrices
$\VEC{\Sigma}_{1} \dots \VEC{\Sigma}_{4}$.}

\label{tab:para}
\begin{center}
\begin{tabular}{c|cccc}\hline\hline
	Parameter  &   value   &  \\
	\hline
	$\zeta_{1}$ & 0.281 \\
	$\zeta_{2}$ & 0.252 \\
	$\zeta_{3}$ & 0.276 \\
    $\zeta_{4}$ & 0.189 \\
	\hline
	$\VEC{\mu}_{1}$	& $\left(\begin{array}{l l l}$-$0.454 & $-$0.497 &
1.276\end{array}\right)$	\\
	$\VEC{\mu}_{2}$	& $\left(\begin{array}{l l l} $-$0.058 & 0.234	& 0.549
\end{array}\right)$	\\
	$\VEC{\mu}_{3}$	& $\left(\begin{array}{l l l} $-$0.310 & $-$0.335	& 1.039
	\end{array}\right)$	\\
    	$\VEC{\mu}_{4}$	& $\left(\begin{array}{l l l} $-$0.552 & $-$0.165	& 1.501
	\end{array}\right)$	\\
	\hline
    
	$\VEC{\Sigma}_{1}$  & $\left(\begin{array}{r r r} 2.04\times 10^{-3} & -1.93 \times 10^{-3} & -2.41\times 10^{-3} \\
    -1.93 \times 10^{-3} &  3.18 \times 10^{-2} & -5.16 \times 10^{-3}\\
    -2.41 \times 10^{-3} & -5.16 \times 10^{-3} &  4.04 \times 10^{-2}\\
    \end{array} \right)$ \\
$\VEC{\Sigma}_{2}$  & $\left(\begin{array}{r r r} 3.65\times 10^{-2} & 1.68 \times 10^{-2} & -4.99\times 10^{-2} \\
    1.68 \times 10^{-2}  & 7.99 \times 10^{-2} & 9.57 \times 10^{-3}\\
    -4.99 \times 10^{-2} & 9.57 \times 10^{-3} & 2.35 \times 10^{-1}\\
    \end{array} \right)$ \\
$\VEC{\Sigma}_{3}$  & $\left(\begin{array}{r r r} 8.42\times 10^{-3} & 1.04 \times 10^{-2} & -1.45\times 10^{-2} \\
     1.04 \times 10^{-2} &  4.98 \times 10^{-2} & -5.07 \times 10^{-2}\\
    -1.45 \times 10^{-2} & -5.07 \times 10^{-2} &  1.21 \times 10^{-1}\\
    \end{array} \right)$ \\
    $\VEC{\Sigma}_{4}$ & $\left(\begin{array}{r r r} 1.90\times 10^{-2} & -2.87 \times 10^{-2} & -1.42\times 10^{-2} \\
    -2.87 \times 10^{-2} &  5.49 \times 10^{-2} & -2.74 \times 10^{-2}\\
    -1.42 \times 10^{-2} &  2.74 \times 10^{-2} &  6.62 \times 10^{-2}\\
    \end{array} \right)$ \\
		\hline \hline
\end{tabular}
\end{center}
\end{table}

\subsection{Internal cluster validation}

Cluster validation plays a key role in assessing the quality of a given clustering structure. It is called internal when statistics are devised to capture the quality of the induced clusters using solely the available data objects. Four validation measures are used: Bayesian information criterion \citep[BIC;][]{Schwarz1978}, but see also \citet{Drton2017},  integrated complete likelihood \citep[ICL;][]{Biernacki:2000}, entropy \citep{Baudry2010}, and silhouette \citep{rousseeuw1987} diagnostics.  See Appendix \ref{ap:diag} for details of each of the former methodologies.  Below,  the  scrutiny of  the adopted diagnostics  for GMM solutions up to 10  GCs.

\begin{figure*}
\includegraphics[width=1\linewidth]{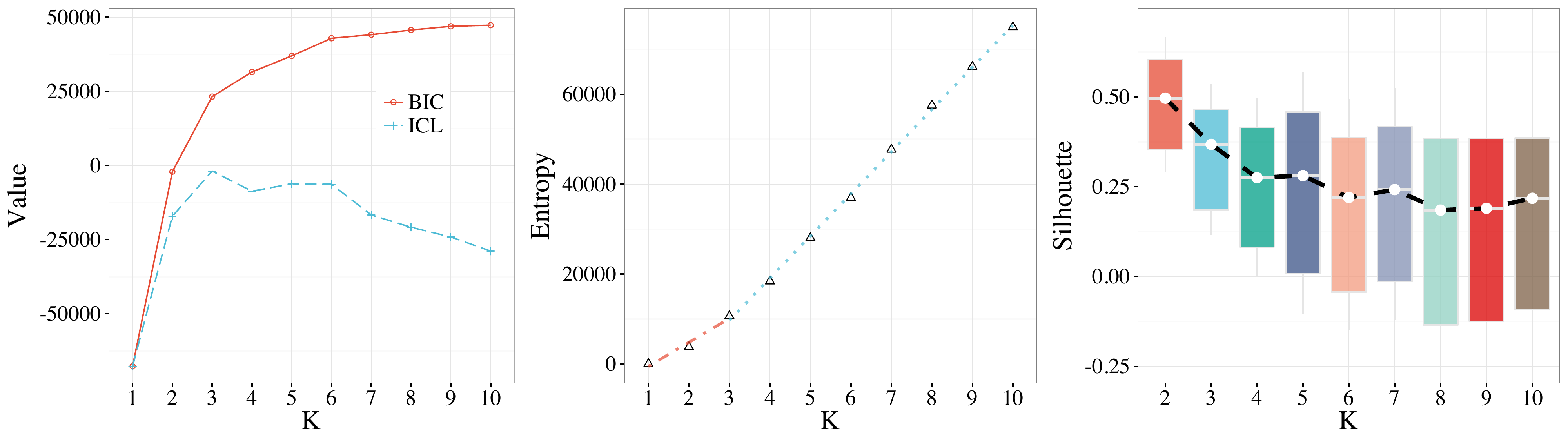}
\caption{Results for $K$ number of Gaussian components for each of the internal validation methods as follows: the leftmost panel shows the BIC and ICL values, the middle panel shows the entropy elbow diagnostics,  and the rightmost panel shows the silhouette results.}
\label{fig:diagnostics}
\end{figure*}

BIC and ICL solutions are shown on the left panel of Figure~\ref{fig:diagnostics}. Note that BIC fails to constrain the model to a reasonably low number of groups.  ICL on the other hand suggests a lower number of GCs. Similar differences between BIC and ICL are well known in the statistical literature \citep[e.g.][]{Biernacki:2000}.  Entropy values for solutions ranging from 2 to 10 clusters are shown in the middle panel of Figure \ref{fig:diagnostics}. There is an elbow in the plot at $K$ = 3 GCs, which, together with the preference of ICL, leads us to focus our attention around this solution. Additionally, the silhouette values are displayed  in the rightmost panel of Figure \ref{fig:diagnostics}, suggesting the preference for only two groups. 

At this point, multiple standard internal validation methods suggest that 2-3 clusters are present in the data, but the discrimination is not clear and more cluster components are compatible with the internal validation measurements.
 In the following,  we propose the use of a residual analysis to \textit{break the tie}  and  quantify how well each model performs in terms of synthetically reproducing the original data structure.

\subsection{Residual analysis}

Residual analysis is one of the most informative  methods to check a model fit. It helps to measure  how well a statistical model explains the data at hand and its ability to predict future sets of observations \citep[see e.g. ][for applications of residual analysis in the context of  mixture models]{Bruce92,Cui2015}.  

\begin{figure*} 
\includegraphics[width=0.475\linewidth]
{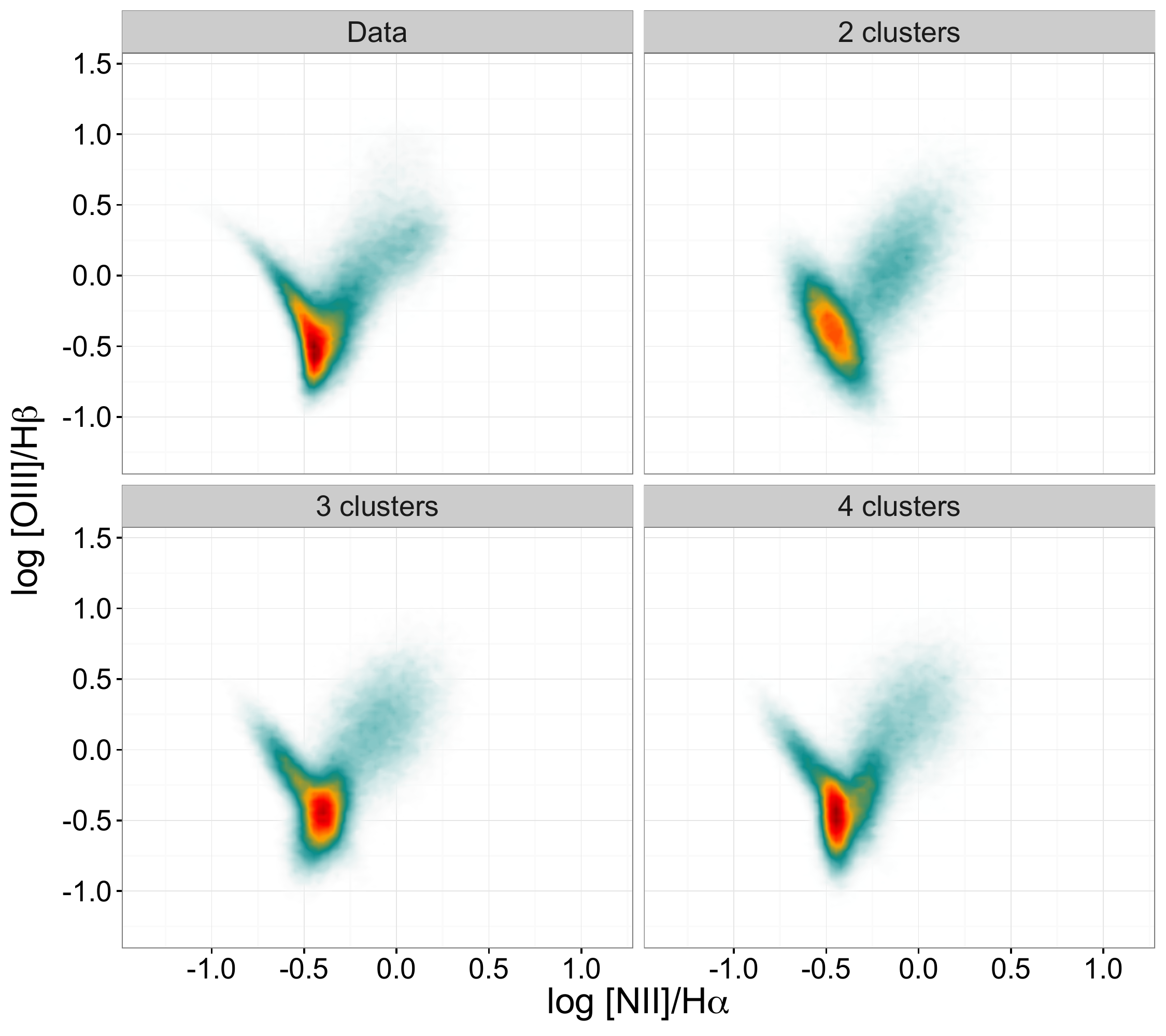}
\includegraphics[width=0.475\linewidth]
{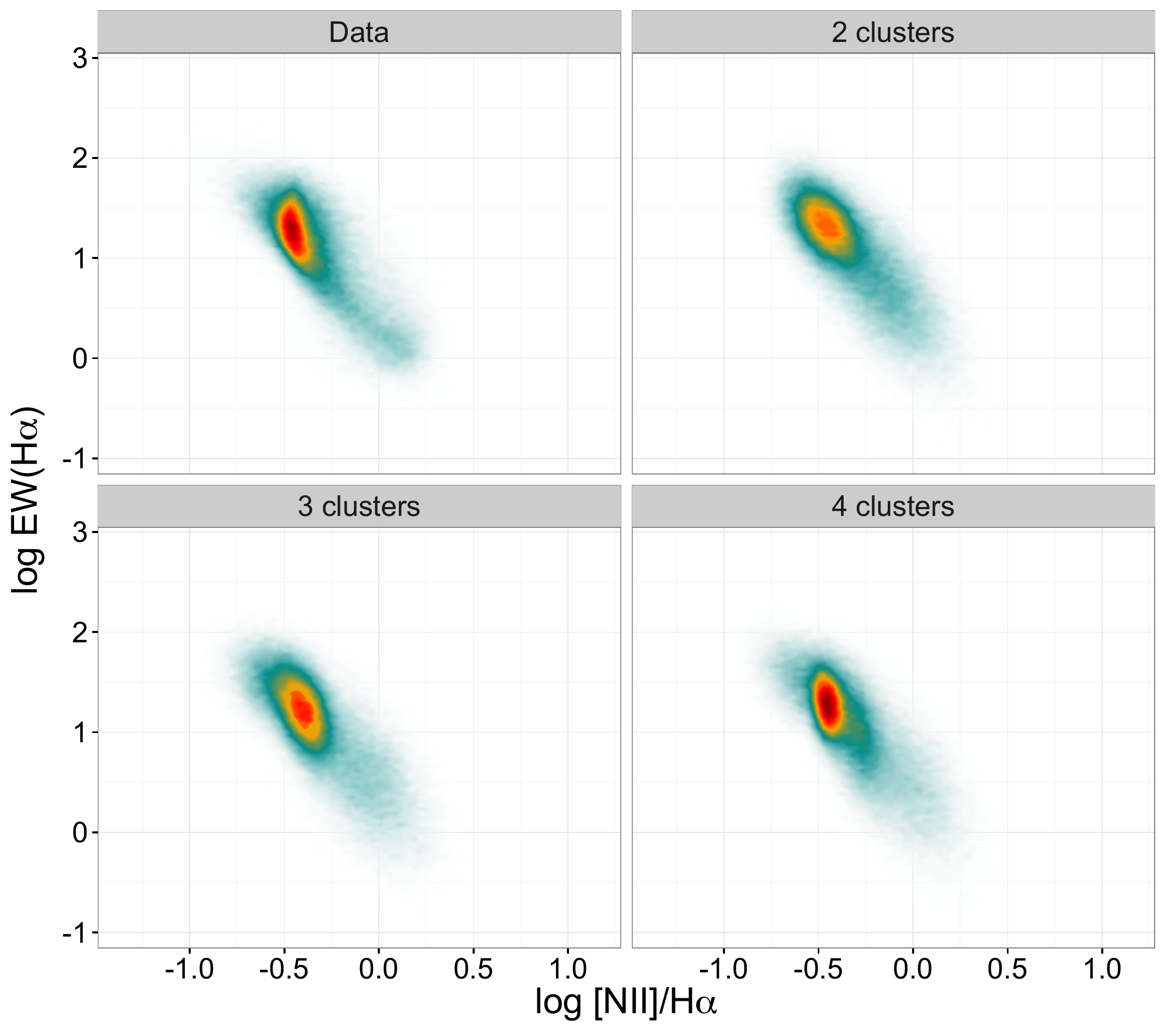}
\caption{Comparison between observed and synthetic data for two-, three-, and four-cluster solutions on the BPT and WHAN diagrams. Both the original and synthetic data are smoothed using the same kernel.}
\label{fig:simulation_BPT}
\end{figure*}

In order to check the goodness of fit of each GMM solution, a comparison between the synthetic and observed data for each model projected onto the BPT and WHAN diagrams is performed. Results are presented in Figure \ref{fig:simulation_BPT},  which shows the smoothed observed  data 
contrasted with the GMM solutions for two, three, and four GCs\footnote{To smooth the residual maps, we use kernels with a bandwidth of 0.05 within a grid of $100 \times 100$.}. A visual 
analysis of Figure \ref{fig:simulation_BPT} reveals that on the BPT diagram, the solution with two clusters barely reproduces the two-wing shape, and the four-cluster solution seems to be a nearly perfect match with the original data. On the WHAN diagram a visual inspection 
does not does not lead  to equally clear conclusions, but the solution with four groups seems to be preferred.  

\begin{figure*} 
 \includegraphics[width=0.485\linewidth,height=0.4\linewidth]
 {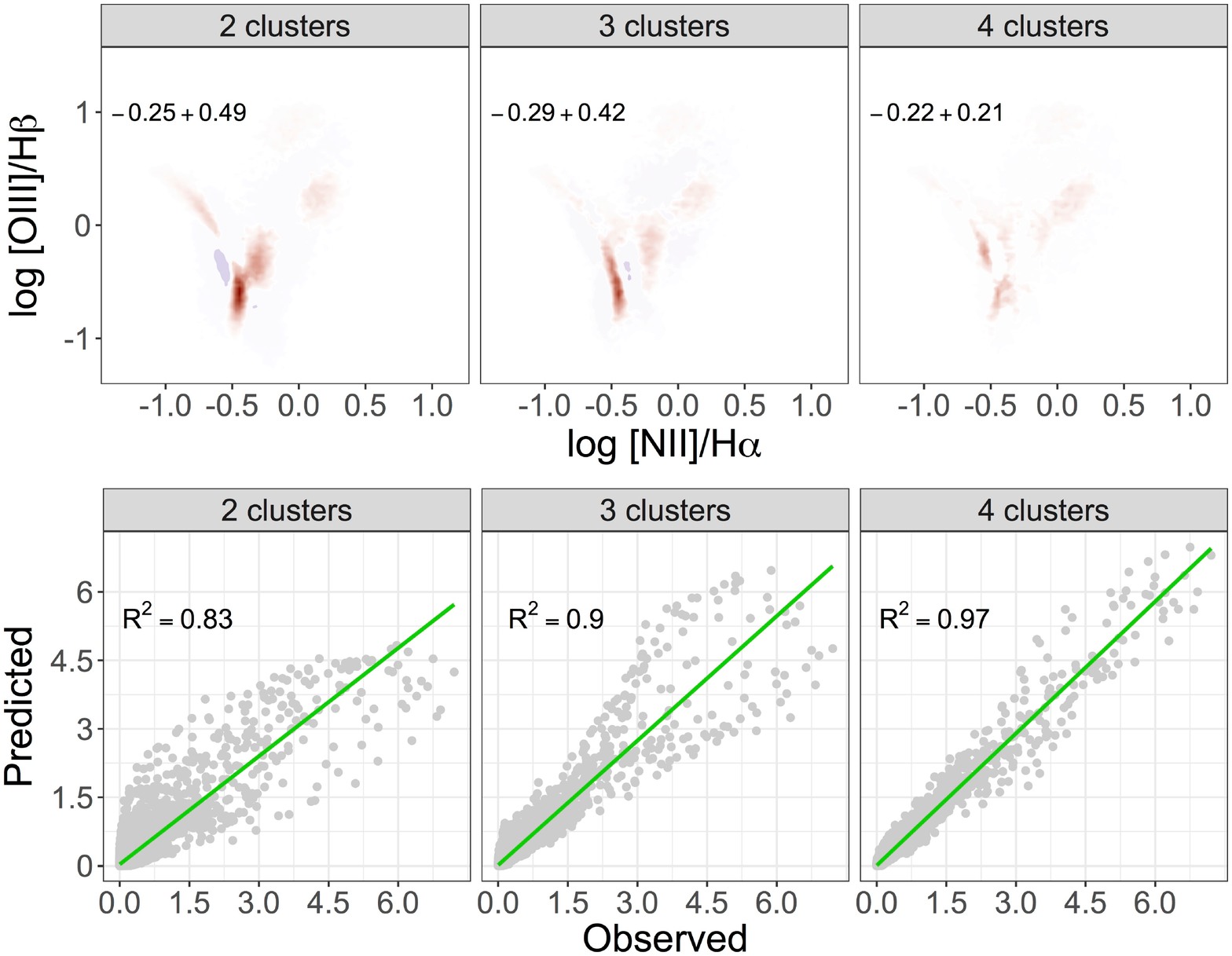}
  \includegraphics[width=0.485\linewidth,height=0.4\linewidth]
 {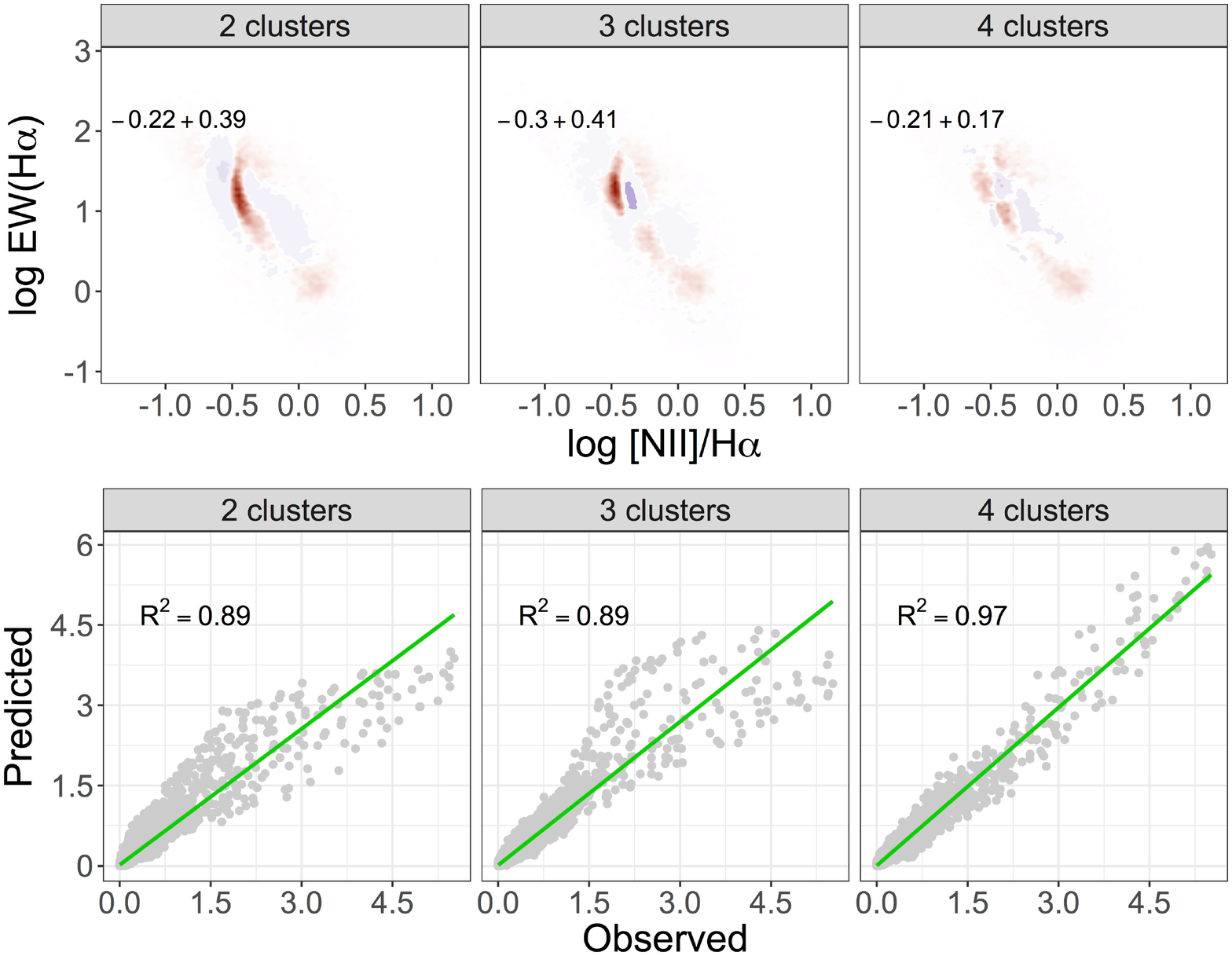}
 \caption{Goodness of fit diagnostics for the 2, 3, and 4 Gaussian components solutions projected onto the BPT (left panel) and WHAN (right panel) diagrams. Top: residual surface density, negative residuals are shown
in blue and positive residuals are shown in red.  Also displayed  is the maximum variation of the residual map in comparison to the original data. Note that for the solution with four GCs, the maximum difference between the  simulated and original data is always $\lesssim 22 \%$.  
Bottom: surface density of the  mixture model solution is plotted
against surface density of the smoothed observed data. A linear fit  of predicted \emph{vs}. observed values is green, and on the left side of each panel we indicate the proportion of variance explained,  $R^2$.}
\label{fig:diag_btp}
 \end{figure*}

A quantitative  analysis of  Figure \ref{fig:simulation_BPT} is displayed in  Figure~\ref{fig:diag_btp}, which shows  the residual map for each solution together with a linear fit between the smoothed observed and simulated data for each GMM  solution.  
Note that for the BPT diagram, each increase in  the number of GCs considerably improves the amount of variance explained by the model, which  is consistent with the visual analysis of Figure \ref{fig:simulation_BPT}. The solution with four GCs is  able to explain up to  $97\%$ of the data variance.  For the WHAN diagram the distinction between the solutions with two and three clusters is fuzzy,  but the solution with four GCs is equally capable of explaining $97\%$ of the data variance.  
Combining the residual analysis and the internal validation methods previously described leads us to keep the solution with four GCs as our "fiducial model" hereafter.

\section{External Cluster Validation Applied to the GMM solution}
\label{sec:cluster_assessment}

This section discusses how to attribute  physical  meaning to the statistically motivated groups, and  provide the means on how to compare them to current classification schemes. If the cluster validation is performed  against an external and independent classification of objects (e.g. the BPT and WHAN classifications), the validation is called external. External cluster validation (ECV) is based on the assumption that an understanding of the output of the clustering algorithm can be achieved by finding a resemblance of the clusters to existing classes. In the present application,  ECV is used to compare the cluster structure produced by a GMM to the class structure corresponding to well-established galaxy classification schemes.

The idea is to use an objective methodology to decide if the induced clusters have recovered an existing classification, or if there is evidence to claim the existence of novel data groups not resembling existing classes. Specifically, we use a probabilistic approach to individually compute 
the distance between each cluster and its most similar class; the degree of separation between cluster and class can then be analysed to decide on the scientific value behind a small distance (near match) or long distance (strong disagreement). The methodology herein employed follows the work of \citet{Vilalta2007}; we briefly describe its  main concepts in appendix \ref{app:cluster_val}. The method relies on the estimate of the \textit{Kullback–Leibler} distance \citep[KL;][a measure of relative entropy]{Kullback1951}  between different groups projected into one dimension via linear discriminate analysis (LDA).  Smaller KL distances are found for closer groups.   

\begin{figure} 
\includegraphics[width=\linewidth]
{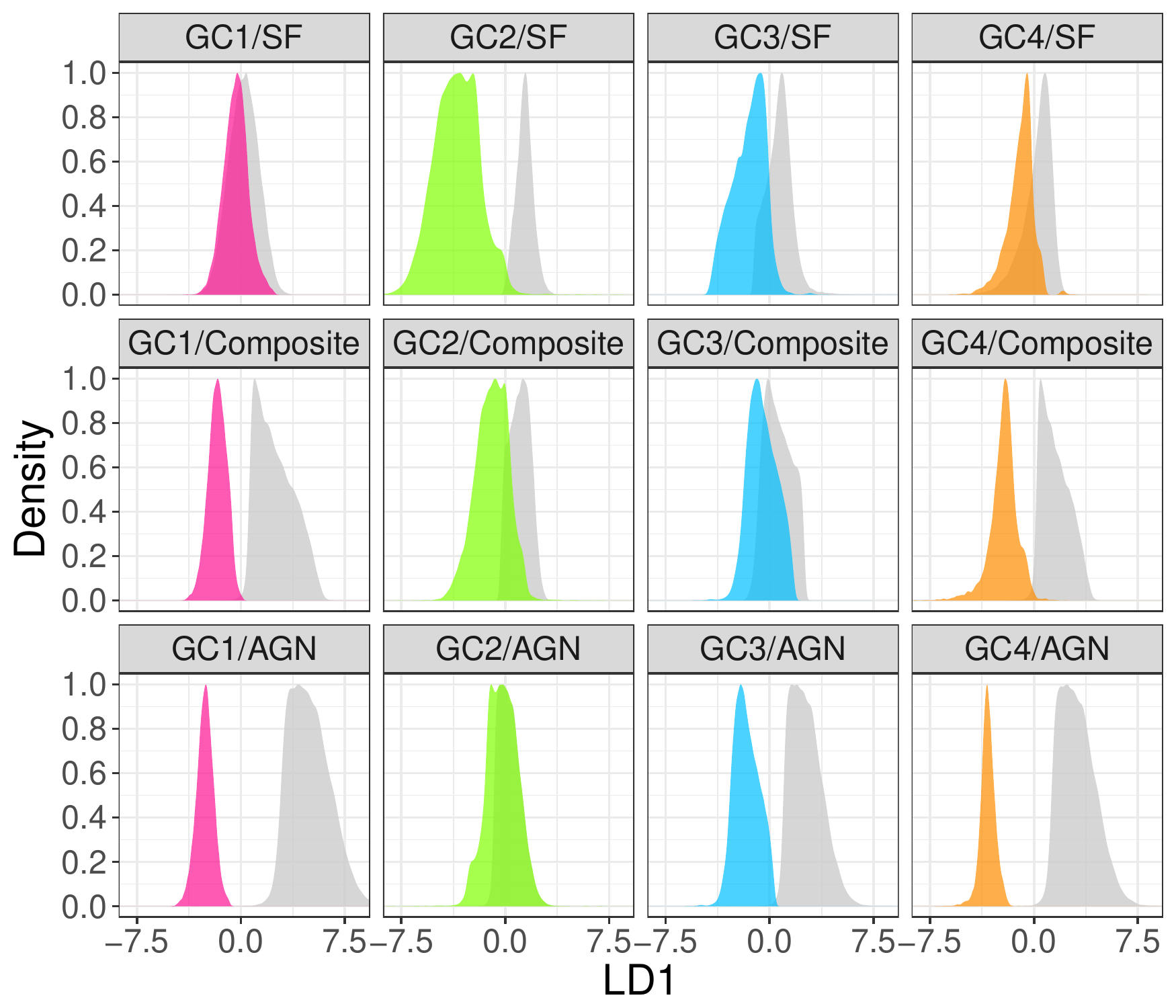}
\caption{Density distributions in the one-dimensional linear discriminant projections for each of the 4 Gaussian components (coloured distributions) compared to the traditional BPT classification of SF, composite, and AGN galaxies (grey distributions).}
\label{fig:mosaic_BPT}
\end{figure}

\begin{figure} 
\includegraphics[width=\linewidth]
{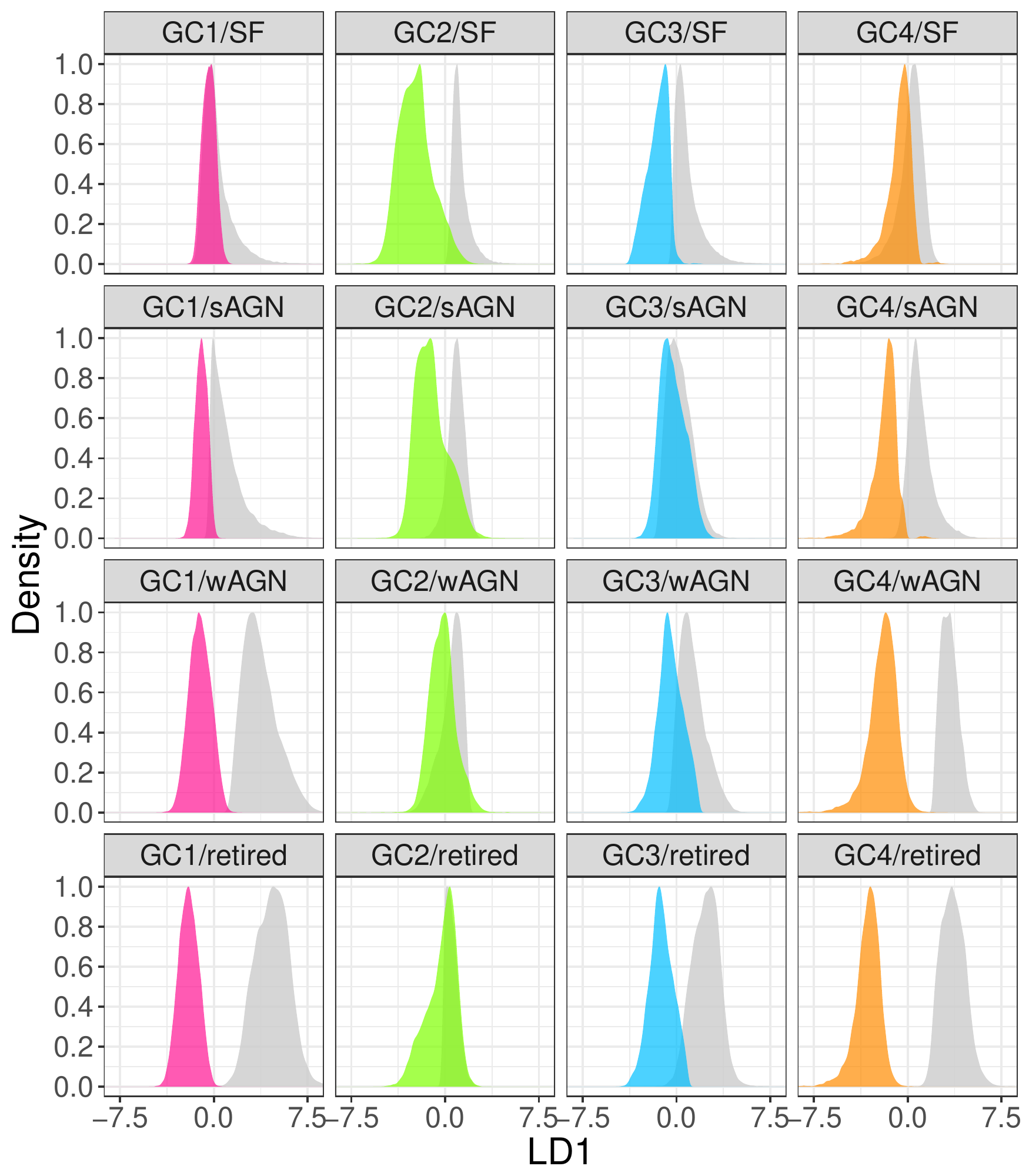}
\caption{Density distributions in the one-dimensional linear discriminant projections for each of the 4 Gaussian components (coloured distributions) compared to the traditional WHAN classification of SF, sAGN, wAGN, and retired galaxies (grey distributions).}
\label{fig:mosaic_whan}
\end{figure}

To illustrate the application of ECV methodology in our dataset, we show a  pairwise comparison of the one-dimensional linear discriminant projections of the four GCs  to the BPT and WHAN classifications in Figures~\ref{fig:mosaic_BPT} and~\ref{fig:mosaic_whan}, respectively. The figure depicts each group -- colour-coded as in Figure ~\ref{res:GMM_2d_BPT_WHAN} -- alongside the astrophysically motivated classes. 

A visual inspection of  Figures~\ref{fig:mosaic_BPT} and~\ref{fig:mosaic_whan} reveals that groups GC1 and GC4 are closer to SF galaxies in both BPT/WHAN diagrams, while GC2 and GC3 are closer to AGN/(wAGN \& retired/passive) and composite/sAGN, respectively. Thus, there is no particular evidence for a new group, but surprisingly GMMs are capable of automatically identifying groups of galaxies resembling the traditional classification scheme of both diagrams from a  higher dimensional feature space. Table \ref{table:comparison} summarizes  the results  showing  each class and  its closest  GC alongside to 1-KL distance. 

\begin{center}
\begin{table}
\caption{Summary of the associations between GMM, BPT and WHAN groups. Next to each class is one minus the KL distance of each BPT and WHAN class to its respective GC.}
\begin{tabular}{ l c c }

GMM & BPT  & WHAN \\
\hline
GC1 & Star forming (0.981) & Star forming (0.996) \\ 
GC2 & AGN (0.981)  & wAGN(0.961) + retired (0.983) \\  
GC3 & Composite (0.943) & sAGN (0.984)   \\
GC4 & Star forming (0.934) & Star forming (0.957)
\end{tabular}
\label{table:comparison}
\end{table}
\end{center}

\begin{figure*}
\includegraphics[width=0.45\linewidth]
{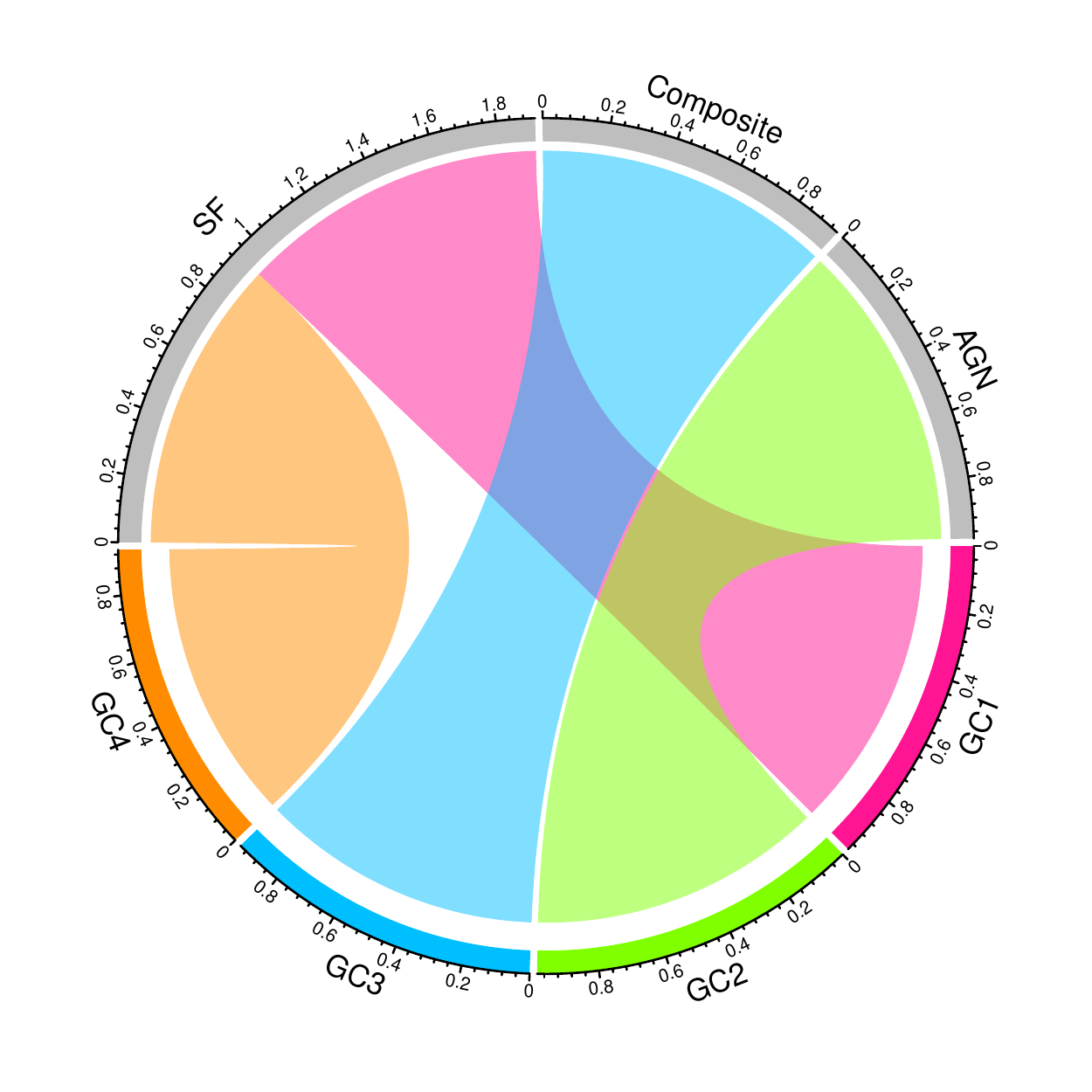}
\includegraphics[width=0.45\linewidth]
{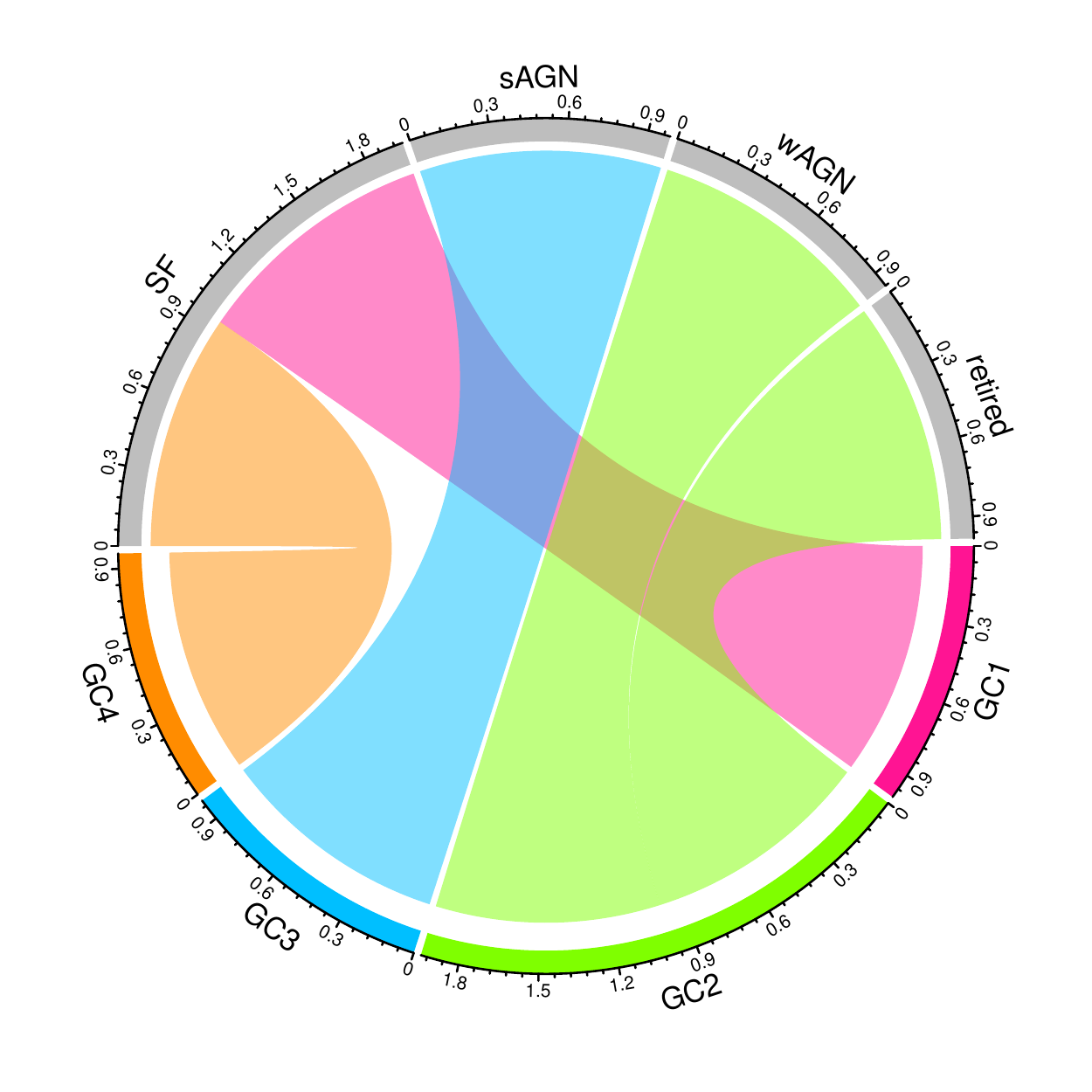}
\caption{Chord diagrams representing the associations between the Gaussian components  and the astronomical classification classes defined on the BPT, left panel, and the WHAN, right panel, diagrams. A thicker connecting ribbon indicates stronger association.}
\label{fig:chord}
\end{figure*}

In order to better visualize the closest  associations (i.e. with a normalized KL $\lesssim 0.05$), we  show a chord diagram \citep{Gu2014,deSouza2015}  in  Figure \ref{fig:chord}. 
It illustrates the level of relationship between distinct groups, which are represented by segments around the circle. Normalized distances between distributions are shown as ribbons; the thickness of the ribbons is weighted by 1-KL distance between each pair of groups, so the thicker the ribbon, the closer the GC to its traditional classification counterpart.

\subsection*{GC1/GC4 Interpretation}

The connection between GC1, GC4 and the SF region in both BPT and WHAN diagrams is straightforward and somewhat expected. In terms of the WHAN diagram, by construction \citep{CidFernandes2011}, the  vertical line at \NHa -0.40 represents an optimal transposition of the \citet{Stasinska2006} SF/AGN-BPT division projected into the WHAN diagram. Consequently, the combination of both diagrams in a three dimensional space should still preserve the same locus for SF dominated galaxies, which is automatically retrieved by the GMM methodology. 

Albeit the solution with 3 GCs (90\% of the data variance) only requires a single GC within the SF region, the solution with 4 GCs (97\% of the data variance) splits the SF region into GC1 and GC4, a behaviour that may be physically interpreted by the presence of starburst galaxies, predominantly populating the top-left wing of the BPT diagram. To be more specific, galaxies at the top-left wing have current specific SFRs about 2 orders of magnitude larger than the metal-rich galaxies at its bottom\citep[see Fig. 2,][]{Asari2007}.

\subsection*{GC2 Interpretation}

The locus occupied by the GC2 in the 3-dimensional emission-line space is concomitantly connected to the BPT-AGN region and mainly wAGN + retired region (seconded by the sAGN region) in the WHAN diagram. The result may appear controversial at first glimpse, as one could expect that the BPT-AGN galaxies should relate to  the sAGN galaxies in the WHAN diagram. Nonetheless, the GMM recovers a previous finding by \citet{Cid2010}. The authors show that the dichotomy between Seyferts and LINERs is wiped out by the presence of weak line galaxies in the sample, which are usually left out from vanilla emission-line galaxy studies solely relying on the BPT plane. They suggest that the right-wing of the BPT diagram is actually populated by AGN and retired galaxies, which is corroborated by the GMM  results. 
In other words, by finding a larger group composed by AGN-BPT and wAGN + retired-WHAN galaxies, our method does not show a statistical evidence for the separation between Seyferts and LINERs as independent sub-classes.

\subsection*{GC3 Interpretation}

As aforementioned, Figures~\ref{fig:mosaic_BPT} and~\ref{fig:mosaic_whan} indicate that GC2 also relates to the sAGN region at the WHAN diagram in a lesser extent than the GC3, which we shall discuss next.   
GC3 relates mostly to the composite-BPT and sAGN-WHAN regions. While it is desirable that GMM finds the composite-BPT locus, the connection to the sAGN-WHAN galaxies is not so straightforward. This can be explained due to the lack of a formal composite area in such diagram. 
From Figure \ref{res:GMM_2d_BPT_WHAN}, we see that GC3 occupies a transitional region between GC1/GC4, and GC2; a locus that could also be designated as an ``effective composite" area between SF and sAGN dominated galaxies.

\section{Age, metallicity,  and 4000 \AA~ break distributions for the GMM Groups} \label{sec:astrophysical_discussion}

We now address whether these data-driven groups bring new insights beyond what is given by established classification schemes. This section discusses characteristics of the galaxies in each GC alongside the BPT and WHAN classes. 

It is well known that different galaxy properties share some sort of \emph{symbiotic} relationship; for instance, D$_n$4000 is closely linked to the characteristics of the stellar population \citep[e.g. average population age, metallicity; as described in][and references therein]{Poggianti1997,Blanton2003,Goto2003PhDThesis,CostaDuarte2013,Stasinska2015,Vazdekis2016}. Hence, the aforementioned features serve as proxies to derive other galaxy properties \citep[e.g. star formation rate][]{Tinsley1980,Zaritsky1993,Poggianti1997,Vazdekis2016}.

\begin{figure}
\includegraphics[trim={5.8cm 1cm 12cm 2.8cm},clip,width=\linewidth]
{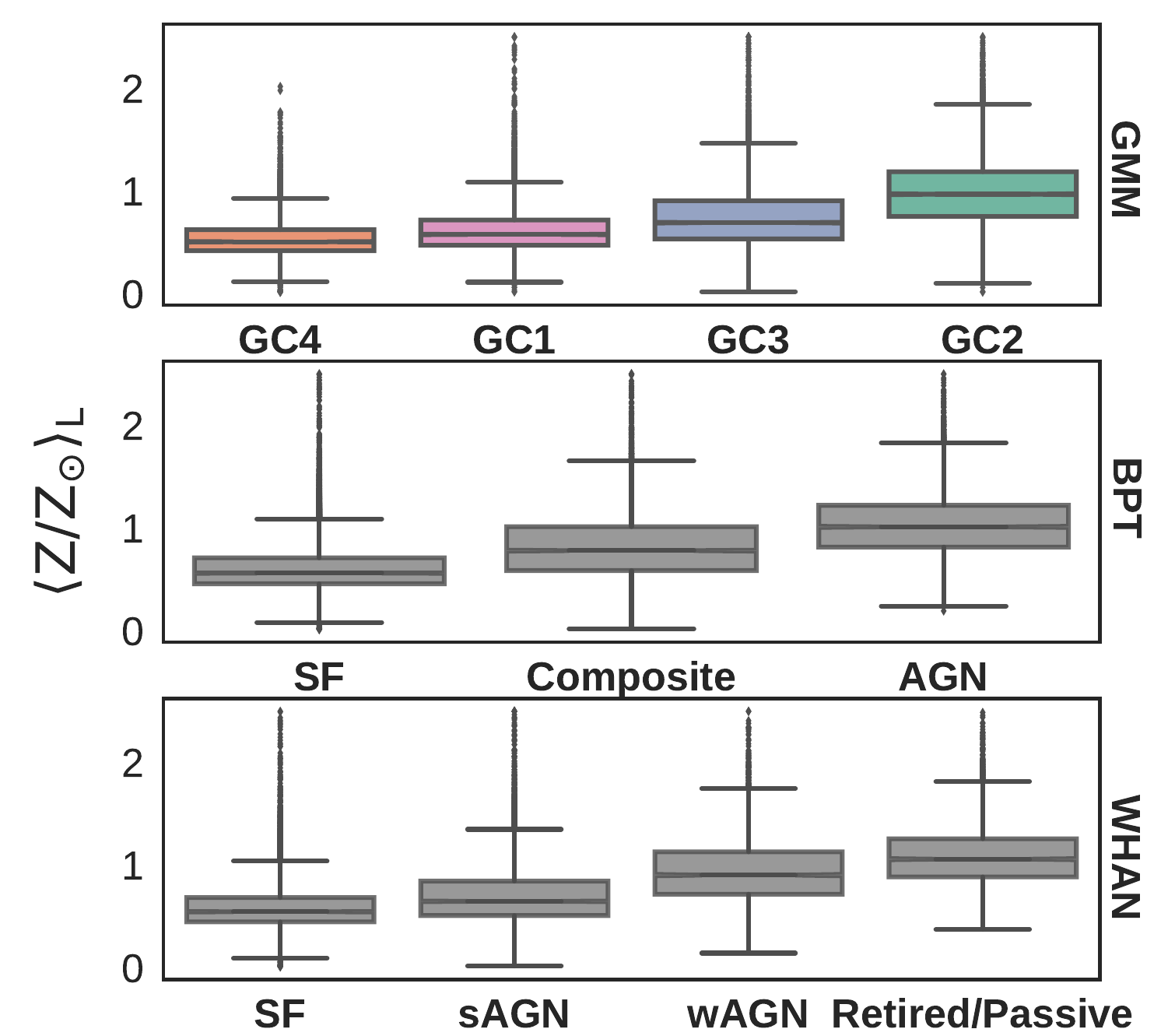}
\caption{Metallicity distributions portrayed as boxplots for GMM, BPT and WHAN diagram classes, from top to bottom. For the GMM we can see the GCs displayed in the following order: GC4, GC1, GC3, GC2 following increasing \NHa, i.e. the $x$-axis of the BPT diagram.
The order of the remaining groups 
is given by the KL distance to the GMM components. The width of boxes is proportional to the square root of the number of galaxies within each bin and the whiskers extend to the most extreme data point, which is within the 50 per cent interquartile range (IQR). 
To better illustrate the overall distribution of the sample, for the boxplots of the BPT diagram we have omitted part of the outlier zone, ``zooming'' into the interquartile distance.}
\label{fig:met_gmm}
\end{figure}

\begin{figure} 
\includegraphics[trim={5.8cm 1cm 12cm 2.8cm},clip,width=\linewidth]
{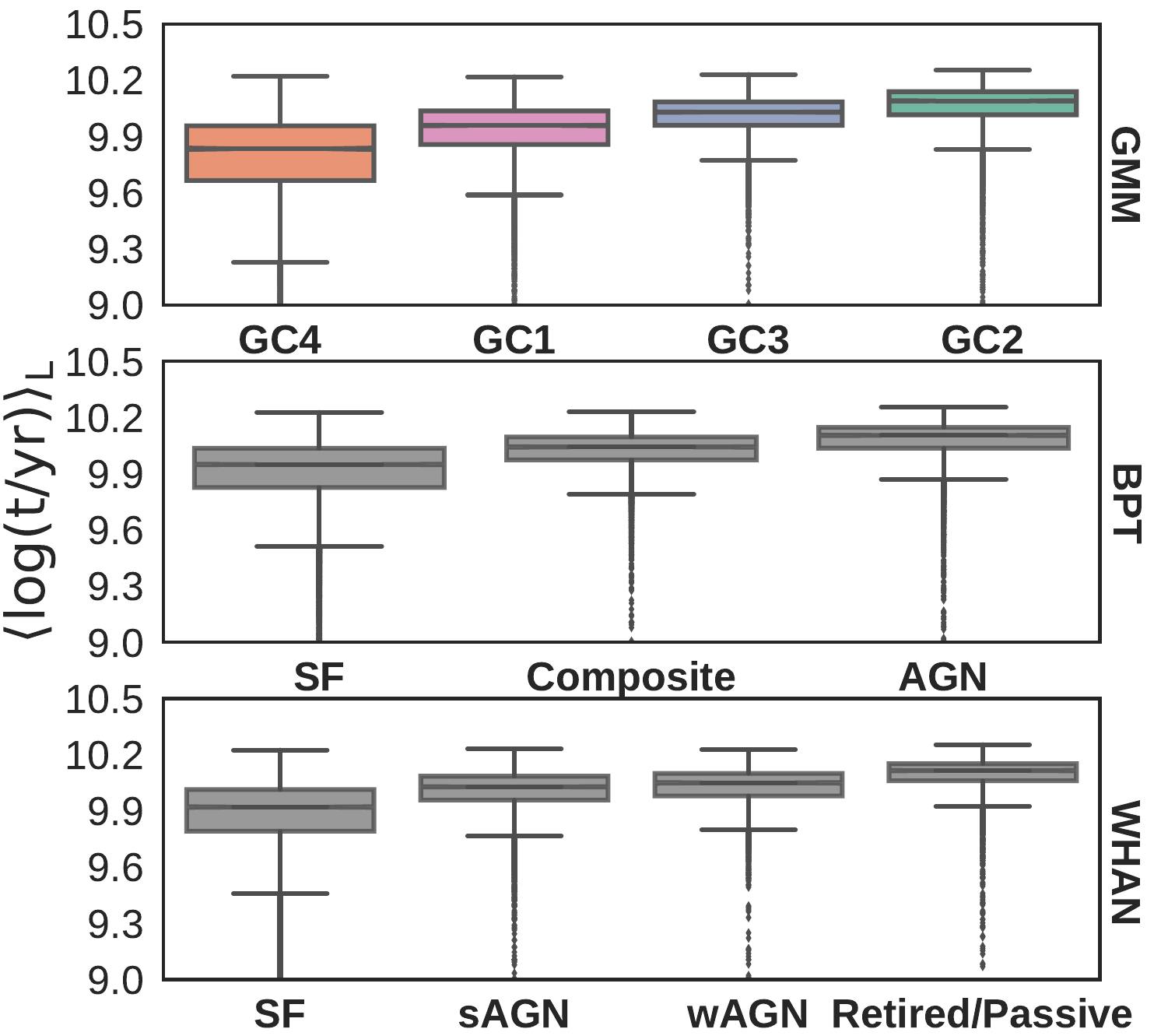}
\caption{Average stellar age distributions portrayed as boxplots for GMM, BPT and WHAN diagram classes, from top to bottom. The GCs and remaining groups are ordered as in Figure \ref{fig:met_gmm}.}
\label{fig:age_gmm}
\end{figure}

\begin{figure} 
\includegraphics[trim={5.8cm 1cm 12cm 2.8cm},clip,width=\linewidth]
{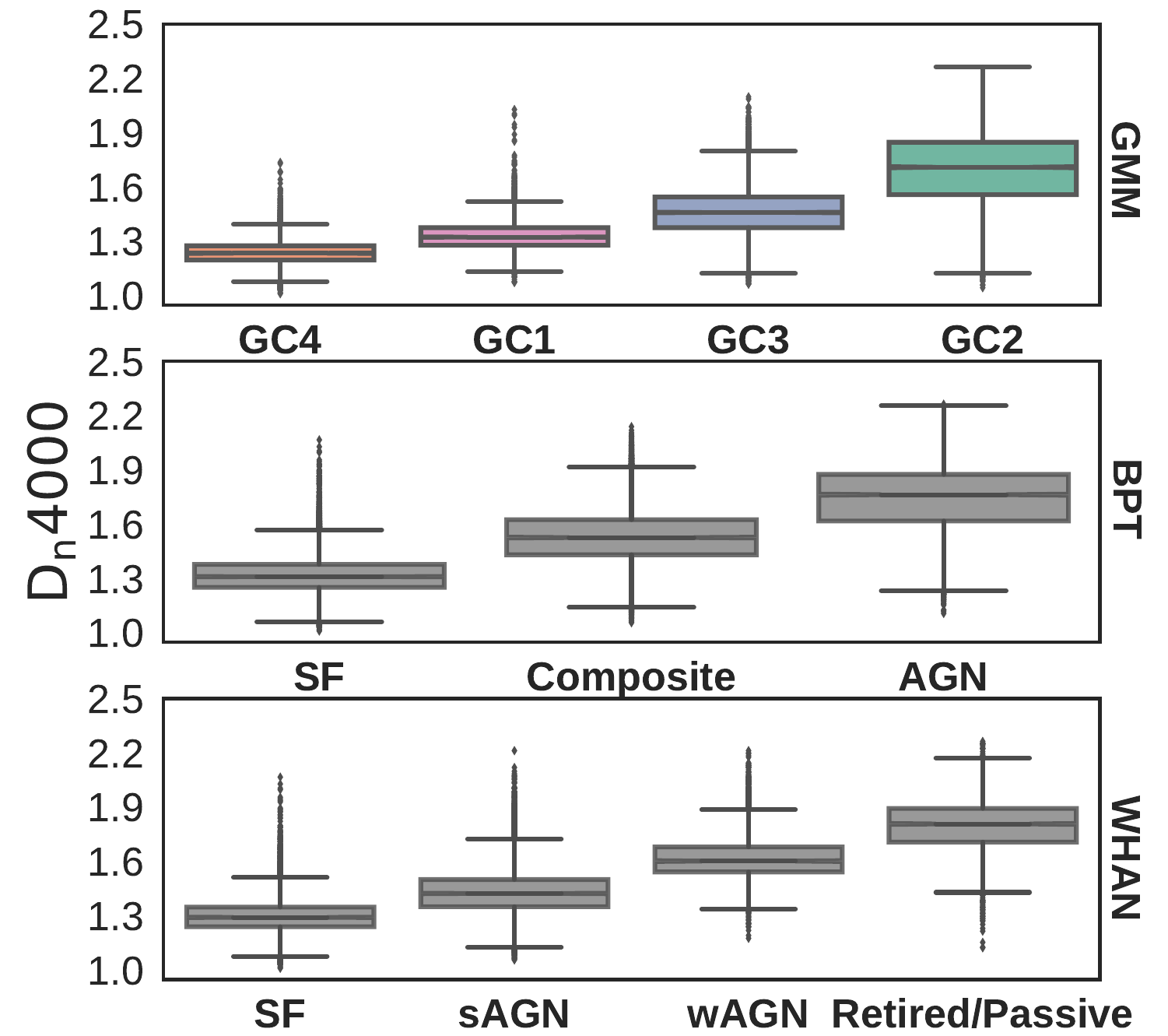}
\caption{D$_n$4000 distributions portrayed as boxplots for GMM, BPT and WHAN diagram classes, from top to bottom. The GCs and remaining groups are ordered as in Figure \ref{fig:met_gmm}.}
\label{fig:d4k_gmm}
\end{figure}

In order to probe how the GCs compare to the classes derived by classical diagrams, we look into their properties not explicitly used in the GMM analysis. For that purpose, we choose three of the main features retrieved from the SEAGal/STARLIGHT output data,  as defined in Section \ref{sec:data}: $\langle Z/Z_{\odot}\rangle_{L}$, $\langle \log (t/{\rm yr})\rangle_{L}$, and D$_n$4000\footnote{Note that we are using average values weighted by flux/luminosity due to the smaller uncertainties on those \citep[see Table 1,][]{CidFernandesetal2005}. In the case of missing values, synthetic D$_n$4000 was used. This is done for a better sampling statistics, but it does not affect the overall results.}.  The goal is to check how these properties vary according to the employed classification: GMM, BPT and WHAN. 
To that end, we portray their statistical properties as boxplots in Figures \ref{fig:met_gmm}, \ref{fig:age_gmm}, and \ref{fig:d4k_gmm}, as well as  their summary statistics in Table \ref{table:properties}, which shows the values for the median, 1\textsuperscript{st} (Q1) and 3\textsuperscript{rd} (Q3) quartiles,  and the interquartile range (IQR $\equiv$ Q3-Q1). 

On the boxplots, the groups are vertically aligned based on their proximity in terms of the KL distance, and the  fiducial order roughly follows increasing values of \NHa\, in the BPT diagram (i.e. from left to right: SF, Composite, AGN). As we can see from the boxplots, by aligning these groups in this way, a positive monotonic relationship between the classes and the median values of  $\langle Z/Z_{\odot}\rangle_{L}$, $\langle \log (t/{\rm yr})\rangle_{L}$, and D$_n$4000 distributions is revealed. The trend is a consequence of the AGN host galaxies having different characteristics from their inactive counterparts, preferentially populating the so-called green valley and red sequence of the colour-mass diagram \citep[e.g.][]{Schawinski2010}. Thus, as we move towards the right side of the BPT diagram, one is mostly looking at early-type galaxies, that are characterized by older and more metallic stellar populations, and higher values of D$_n$4000 \citep[e.g.][]{Poggianti1997,Schawinski2010,DeSouza2016}. 

Notably, the GMM solutions automatically find groups that share meaningful physical properties, beyond the features  used in the clustering algorithm. An inspection of Table \ref{table:properties} confirms that the statistical properties are quite similar between the GMM groups and their respective counterparts on the BPT and WHAN diagrams. Additionally, Table \ref{table:properties} shows that the distributions of galaxy properties in SF groups are consistent between the BPT and WHAN diagrams in terms of medians and IQR. Their values roughly lie between those of GC4 and GC1. For instance, GC4 and GC1 have median values for $\langle Z/Z_{\odot}\rangle_{L}$ of 0.50 and 0.58, while the BPT and WHAN SF groups have values of 0.56 and 0.55, respectively. 
 
In the case of $\langle Z/Z_{\odot}\rangle_{L}$ and D$_n$4000, the median and IQR increase more steadily for the GCs, in comparison to the BPT and WHAN classes. 
The trend of D$_n$4000 visible in Figure \ref{fig:d4k_gmm} indicates that different types of galaxies occupy different loci in the GMM classification. For instance, GC4, the first one on the left, has  low median values for those parameters, which is in agreement with the characteristics of young stellar populations, i.e. SF galaxies. 
On the other hand, GC2, composed mostly of AGN hosts and retired/passive objects, has a  higher median value of  D$_n$4000, which is in accordance with older stellar populations. As D$_n$4000 can be used as  proxy for morphology \citep{Dressler1990,Brinchmann2004}, the higher D$_n$4000 median highlights the AGN preference to reside in early-type galaxies \citep[][]{Schawinski2010,DeSouza2016}. Besides, the larger IQR for GC2, corroborates with the fact that AGN can reside either within early or late-type galaxies. The decreasing trend in terms of median and IQR found for $\langle \log (t/{\rm yr}) \rangle_{L}$, within the GMM groups is overall consistent with the traditional diagrams as well. 

The GMM systematically finds groups that have a sharper differentiation of their values of $\langle Z/Z_{\odot}\rangle_{L}$, $\langle\log (t/{\rm yr})\rangle_{L}$, and D$_n$4000,  when compared to the standard classification, especially  in terms of distributing most of the dispersion into fewer clusters (usually only one), yielding to  a majority of lower-dispersion classes.
These findings elucidate the power of the proposed method -- since the physical parameters were not included in the GMM classification, their favourable behaviour within and between the GCs has been inherently caused by the method applied.

\begin{center}

    \begin{table*}
    \caption{Summary statistics for the $\langle Z/Z_{\odot} \rangle _{L}$, $\langle \log (t/{\rm yr}) \rangle_{L}$, and D$_n$4000 galaxy properties. Shown are the median, IQR, and the first and third quartile for the properties of the GMM, BPT and WHAN groups.}
    \begin{tabular}{ r | c  c  c  c | c  c  c  c | c  c  c  c }
     \multicolumn{1}{c}{\Large{Method}} & \multicolumn{4}{c}{\Large{$\langle Z/Z_{\odot} \rangle _{L}$}} & \multicolumn{4}{c}{\Large{$\langle \log (t/{\rm yr}) \rangle_{L}$}} & \multicolumn{4}{c}{\Large{D$_n$4000}}\\
    \hline
    \hline
    Classification & Median & IQR & Q1 & Q3 & Median & IQR & Q1 & Q3 & Median & IQR & Q1 & Q3 \\ 
    \hline 
    \multicolumn{1}{l|}{\textbf{GMM}} & & & & & & & & & & & & \\
    GC4 & 0.50 & 0.20 & 0.42 & 0.62 & 9.83  & 0.30 & 9.66  & 9.96  & 1.23 & 0.08 & 1.20 & 1.28 \\
    GC1 & 0.58 & 0.25 & 0.47 & 0.72 & 9.96  & 0.18 & 9.86  & 10.04 & 1.32 & 0.10 & 1.28 & 1.38 \\
    GC3 & 0.69 & 0.37 & 0.53 & 0.90 & 10.03 & 0.13 & 9.96  & 10.09 & 1.46 & 0.16 & 1.38 & 1.54 \\
    GC2 & 0.97 & 0.44 & 0.75 & 1.19 & 10.09 & 0.12 & 10.02 & 10.14 & 1.71 & 0.29 & 1.56 & 1.85 \\
    \cline{2-13}
    \multicolumn{1}{l|}{\textbf{BPT}}  & & & & & & & & & & & & \\
    SF         & 0.56 & 0.25 & 0.46 & 0.71 & 9.95  & 0.21 & 9.83  & 10.04 & 1.31 & 0.13 & 1.25 & 1.38 \\
    Composite  & 0.78 & 0.42 & 0.59	& 1.01 & 10.04 & 0.13 & 9.97  & 10.10 & 1.53 & 0.20 & 1.43 & 1.63 \\
    AGN        & 1.01 & 0.40 & 0.82 & 1.22 & 10.10 & 0.11 & 10.04 & 10.15 & 1.73 & 0.27 & 1.62 & 1.89 \\
    \cline{2-13}
    \multicolumn{1}{l|}{\textbf{WHAN}} & & & & & & & & & & & & \\
    SF               & 0.55 & 0.24 & 0.45 & 0.69 & 9.92  & 0.25 & 9.79  & 10.04 & 1.29 & 0.11 & 1.24 & 1.35 \\
    sAGN             & 0.65 & 0.34 & 0.51 & 0.85 & 10.03 & 0.13 & 9.96  & 10.09 & 1.42 & 0.15 & 1.35 & 1.50 \\
    wAGN             & 0.91 & 0.41 & 0.72 & 1.13 & 10.05 & 0.12 & 9.98  & 10.10 & 1.60 & 0.14 & 1.54 & 1.68 \\
    Retired/Passive  & 1.06 & 0.37 & 0.89 & 1.26 & 10.12 & 0.09 & 10.06 & 10.15 & 1.81 & 0.18 & 1.71 & 1.89 \\ 
    \hline      
   \end{tabular}
   \label{table:properties}
   \end{table*} 
      
\end{center}

\section{Digging deeper -- the Seyfert/LINERs dichotomy and the quest for Passive Galaxies}

Internal validation methods present a trade-off between predictive power and simplicity. In other words, a good model should describe the data as best as possible with the fewer number of groups necessary. Whilst our fiducial model based on  diverse criteria  points for a solution  around 3-4 groups, there is a physical motivation to look further and see  if we can spot the  presence of LINERs and discriminate the Passive/Retired galaxies in our sample. 

\begin{figure} 
\includegraphics[width=0.95\linewidth]
{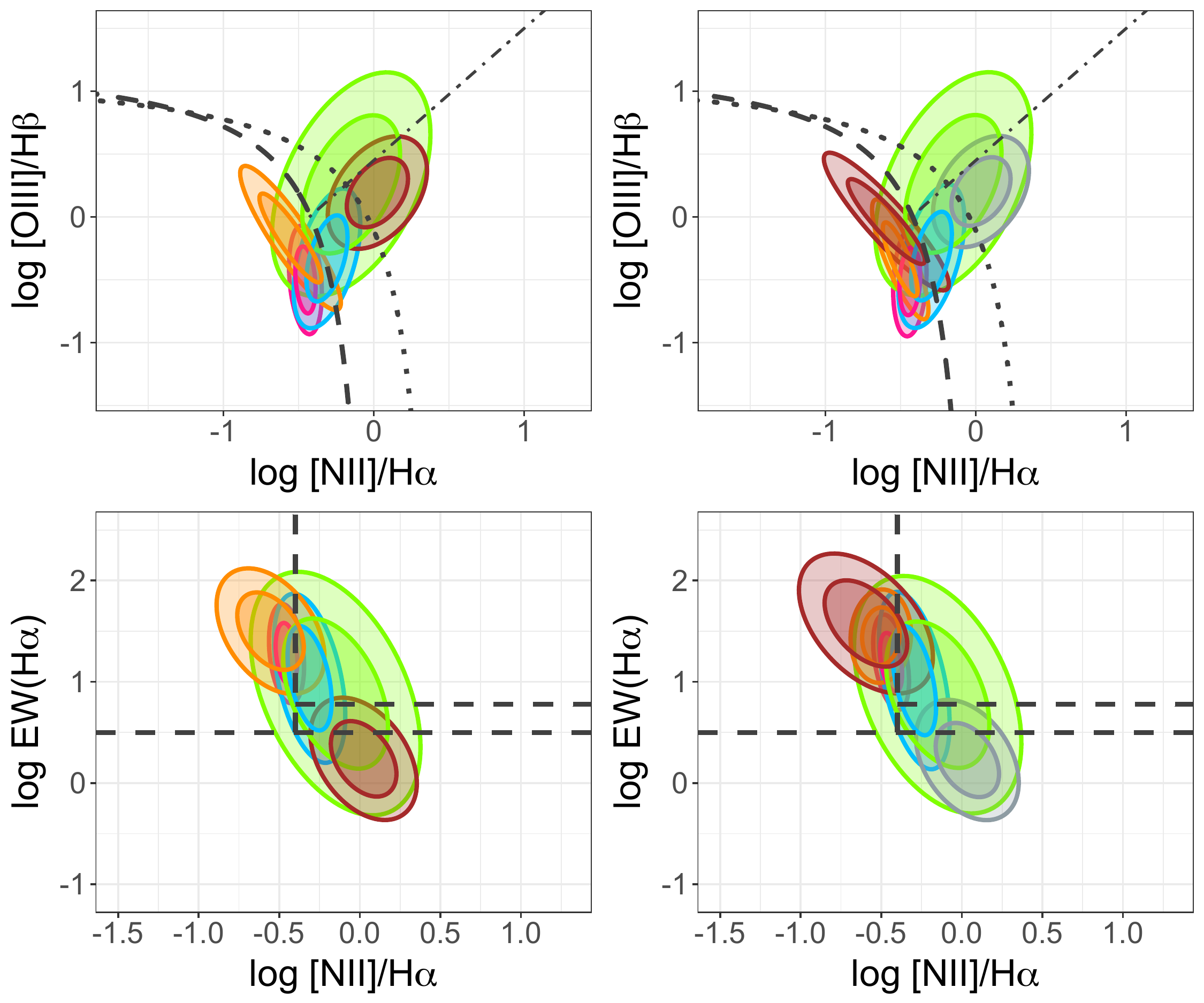}
\caption{The Gaussian components  projected onto the BPT (top panels) and WHAN (bottom panels) diagrams. From left to right are the solutions for 5,  and 6 GCs. For each component the thick lines represent 68\% and 95\%  confidence levels, respectively.
}
\label{res:GMM_2d_BPT_WHAN_56}
\end{figure}

\begin{figure*}
\includegraphics[width=0.45\linewidth]
{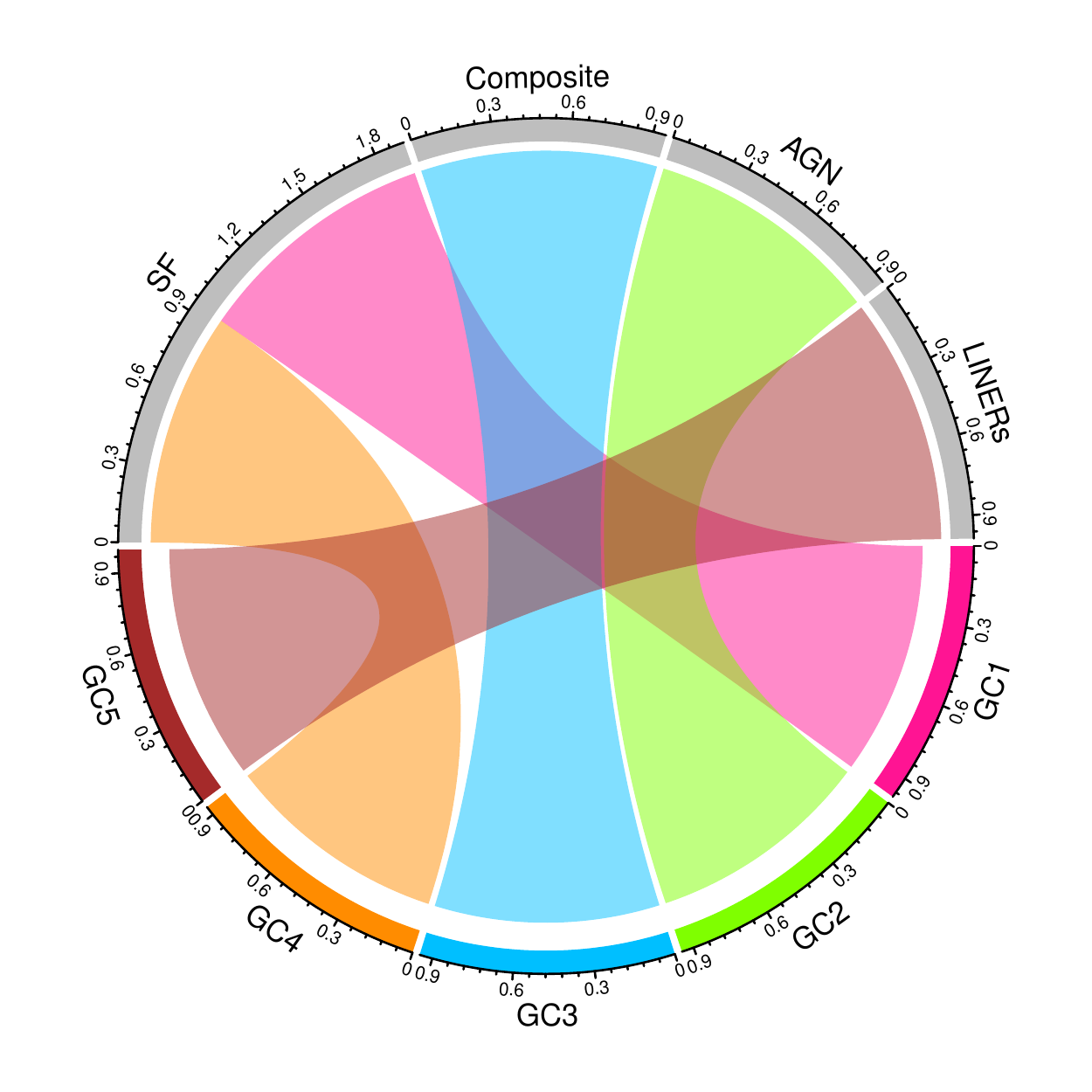}
\includegraphics[width=0.45\linewidth]
{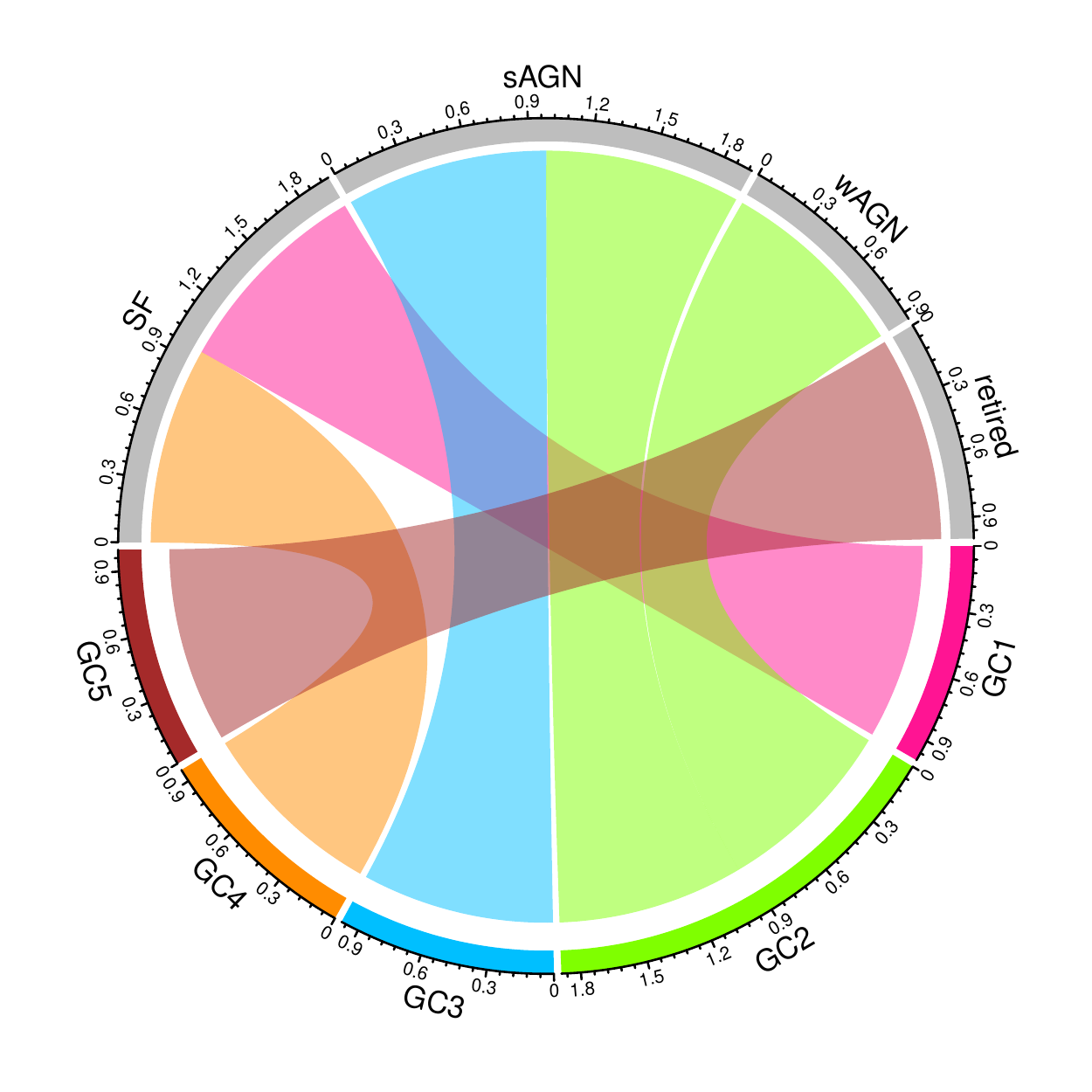}
\caption{Chord diagrams representing the associations between five Gaussian components  and the astronomical classification classes defined on the BPT (with the additional LINER/Seyfert division), left panel, and the WHAN, right panel, diagrams. A thicker connecting ribbon indicates stronger association.}
\label{fig:chord2}
\end{figure*}

\begin{figure*} 
 \includegraphics[width=0.485\linewidth,height=0.4\linewidth]
 {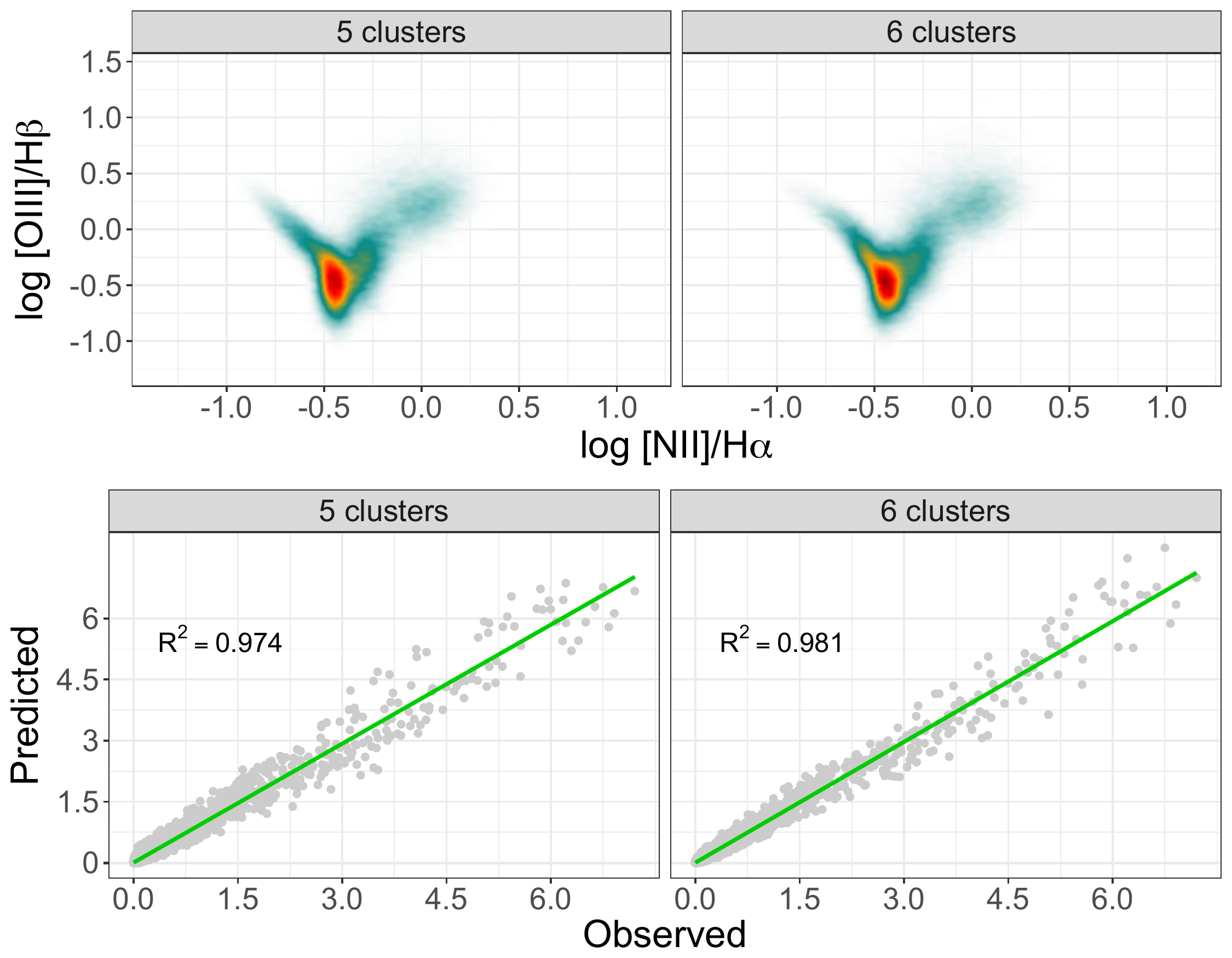}
  \includegraphics[width=0.485\linewidth,height=0.4\linewidth]
 {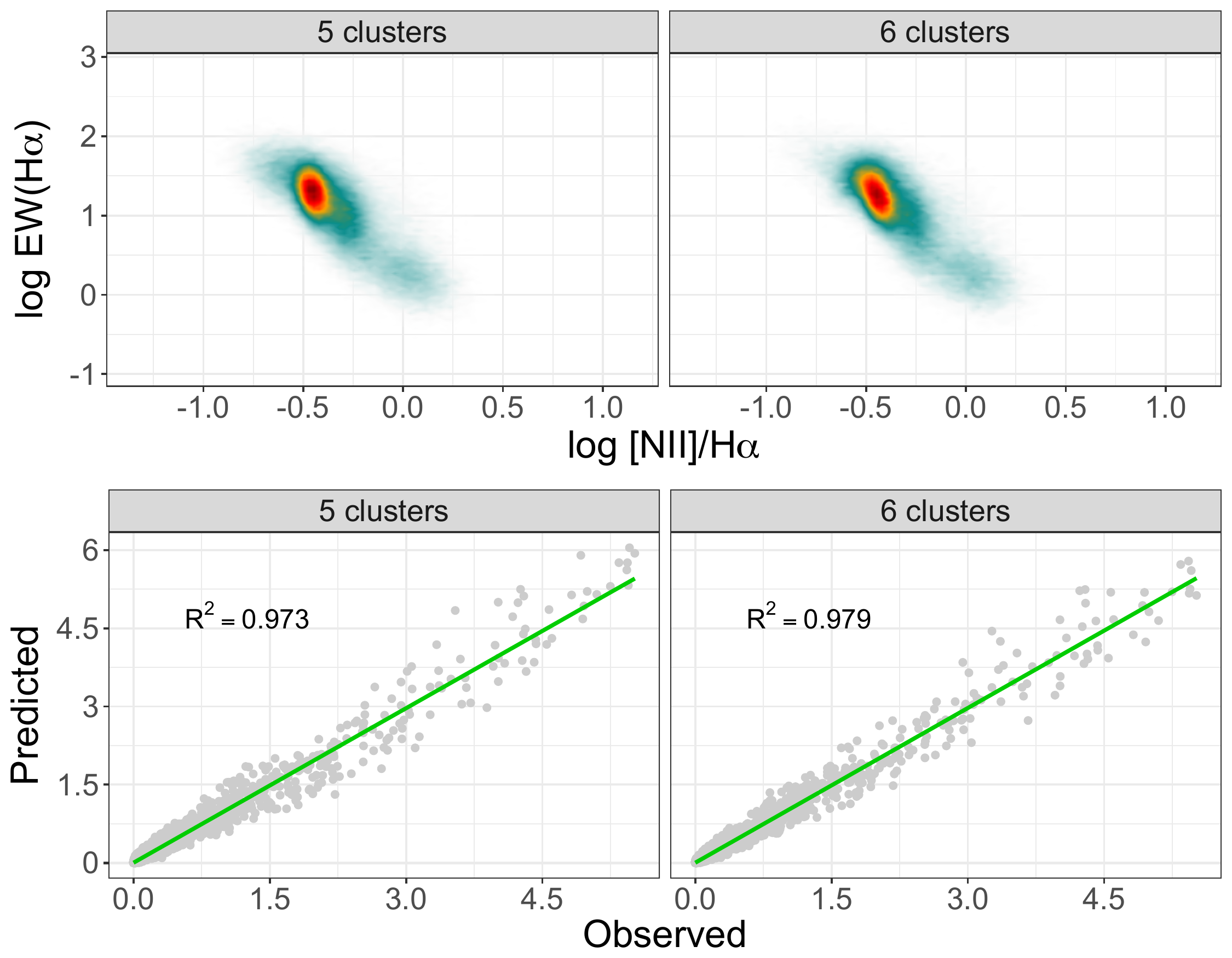}
 \caption{Goodness of fit diagnostics for the 5 and 6 GCs  projected onto the BPT (left panel) and WHAN (right panel) diagrams. Top: Smoothed synthetic data  on the BPT and WHAN diagrams  as in Fig. \ref{fig:simulation_BPT}. 
Bottom: surface density of the  mixture model solution is plotted
against surface density of the smoothed observed data. A linear fit  of predicted \emph{vs}. observed values is green, and on the left side of each panel we indicate the proportion of variance explained,  $R^2$.}
\label{fig:diag_btp_56}
 \end{figure*}

The results of the GMM fit with 5 and 6 GCs are displayed at Fig. \ref{res:GMM_2d_BPT_WHAN_56}, and the corresponding associations, for the solution with 5 GCs, with the BPT and WHAN classification at Fig. \ref{fig:chord2}.
For visualization purposes, the solution with 6 GCs is also shown, but it fragments the SF region into 3 parts, which is mostly driven by its banana-shape rather than by some physical reason. 
The inclusion of GC5 reveals the presence of the LINERs in our sample. As expected, the group also appears connected to the Passive/Retired galaxies class in the WHAN diagram. 
Conversely, the residual analysis depicted at Fig. \ref{fig:diag_btp_56} show that the inclusion of an extra 5 and 6 component increases the level of variance explained, as one should expect from a maximum likelihood estimator for more complex models, but not significantly. Despite the existence of a physical motivation for the use of an extra group, it does not play a major role in explaining the global data variance on this particular feature space. This suggests that a different choice of feature space or the inclusion of an extra-dimension (i.e. emission-lines) could be desirable to make the between-group divisions clearer. While our method is capable of automatically recovering groups that resemble previous classifications and provides the means to evaluate their uncertainties, it does not exclude the importance of the domain expert knowledge in order  to attribute astrophysical meaning to the results. 

\section{Conclusions}
\label{sec:results_discussion}

In this work we develop a data-driven probabilistic approach to classify galaxies, according to their ionization sources, in a three-dimensional space composed of the \OHb, \NHa, and  \EWH \, emission-lines, which represent a joint BPT-WHAN diagram, using public data from SDSS and the SEAGal/Starlight project.  

The results from the parametric Gaussian mixture model are  combined with cutting-edge cluster validation methods, also known as internal cluster validation techniques: BIC, ICL, entropy, silhouette, and residual analysis. This comprehensive study suggests the existence of 4 different classes of galaxies, which are capable to explain up to 97\% of the data variance in both diagrams.  

Given the solution with four groups, an external cluster validation approach is employed to compare the GMM results with previous classification schemes based on domain-expert knowledge (i.e. traditional astronomical classes). The results are visualized using an ubiquitous visualization tool in genetics, also known as chord diagram. 

Our main scientific results and caveats can be summarized as follows:

\begin{enumerate}

\item The best solution for the GMM, based on maximum likelihood estimation and various quantitative evaluation criteria,  has four clusters.  The GMM  statistically retrieves the existence of the  SF, Composite, and AGN BPT-based groups; and the  SF, wAGN, sAGN, and retired/passive WHAN-based groups. 

\item A combination of the GMM results with the external cluster validation technique provides the means to quantify the closeness of each group to their respective counterparts. The SF region (in both diagrams) is divided in two Gaussian components, which  might be a consequence of the  existence of starburst galaxies populating the top-left wing of the BPT diagram. The composite-BPT region and the sAGN region are both connected to the same GMM-based group, which is mostly due to the lack of a formal composite region in the WHAN diagram. The GMM solution indicates the presence of composite galaxies on the WHAN diagram, in an intermediate region comprising part of the traditional SF and sAGN areas. The wAGN+retired/passive galaxies and the AGN-BPT galaxies are connected to a single GMM group as well, which can be explained by the presence of weak line galaxies that populate the right-wing BPT diagram together with AGN-host galaxies. 

\item Within the boundaries of the GMM, and in the three-dimensional optical emission-line feature space where we perform the clustering, our data-driven approach does not find a strong statistical evidence for separate LINER and Seyfert sub-classes for the fiducial solution.  However, LINERs do emerge as a statistically insignificant subgroup when 5 GCs are considered.

\item  To further explore the features of the GMM-based groups against other physical galaxy parameters,  not included in the clustering analysis, we compare the statistical distributions of the $ \langle Z/Z_{\odot} \rangle _{L}$, $\langle \log (t/{\rm yr}) \rangle_{L}$, and D$_n$4000  for the GMM, BPT and WHAN classifications. The GMM  groups have similar statistical properties compared to the standard diagrams, but with a steeper monotonicity in terms of galactic evolutionary properties.  
\end{enumerate}

\subsection*{Our statistical analysis has some limitations and caveats; we address them as follows.}

 \begin{description}

\item[Sample selection;] we decided to work with a volume-limited sample to mitigate the Malmquist bias \citep[e.g.][]{Sandage2000} towards objects with stronger emission lines. If, on the other hand, a magnitude limited samples was chosen, it would include more fainter/dwarf objects. These objects present larger specific star formation and gas fraction galaxies, and preferentially populate the left-wing region of the BPT diagram. It does not change the overall conclusions regarding the number and location of the GCs. The choice of samples slightly affects the location of the fourth Gaussian component responsible for the left-wing region.   

\item[Number of  clusters;]whilst this fiducial model indicates the presence of four GCs, the results should be informed by astrophysical considerations (e.g. photoinization models). The ICL criterion, which is a regularized version of BIC suggests 3 groups as optimal case, with the drawback of explaining only up to 90\% of the data variance, in contrast to the 97\% explained by the use of 4 GCs. A possible solution to explain the extra variance, while still keeping 3 groups would be to use a distorted Gaussian mixture model based for instance in a banana-shape \citep[see e.g.][for an example of how to sample from a banana-shaped distribution]{laine2008adaptive},  or to use a non-linear transformation of the feature space \citep[see e.g.][as an example of how a non-linear coordinate transformation maps a banana-shaped distributions into a  Gaussian one]{Andrew2012}. We should reinforce that the aim here  was to build the best model without compromising simplicity and interpretation. Hence, the methodology is a trade-off between predictive power and parsimony, so we prefer to preserve the original space of features and refrained from use non-parametric models (e.g. DBSCAN, k-nearest) or multi-parametric distributions as e.g. t-mixture models \citep{EMMIXuskew}. Nonetheless, if physically motivated, a more tailored distribution or feature space should be pursued.

\item[Why Gaussian?] We may ask ourselves if  a GMM  is, in fact, a good approximation to explain the data structure of the BPT-WHAN combined subspaces, specially due to the banana-shape of the BPT left wing.  One could apply a more flexible non-parametric method, such as DBSCAN, k-nearest neighbours); or to project the data into a non-linear subspace via e.g. kernel principal components analysis \citep{Ishida2013} or isomaps \citep{Wang2011}. However, in any of these options an important feature would be missing -- again, simplicity and interpretation; which is a hard compromise to get in general in the machine learning approaches.  
The GMM may not be the best possible stochastic model to describe the data, but it is a good trade-off between a parsimonious versus an over-complex model.

\item[Possible follow-ups:] i) the incorporation of physical priors in some based on some flavour of semi-supervised technique, in which information regarding the expected number of groups and their locus could be incorporated and refined; ii) inclusion of additional astrophysical motivated  features. For instance, the  use of the FWHM of [OIII], could unravel extra groups such as the shock-dominated population, since merges are known to  leave imprints in the emission line signal \citep{Leslie2014}; iii) work directly in the raw spectra using a combination of a manifold and deep learning approaches \citep{Sasdelli2016MNRAS} to extract the main spectral features instead of a pre-selected set of emission-line; iv) a comprehensive search for the best lower dimensional subspace able to maximize the discrimination between different galaxy classes.

\end{description}

The analysis herein employed suggests that galaxies with different levels of star formation, with and without supermassive black hole accretion, can be explained by a few classes in low-dimensional spaces. These  classes  have a measurable mean and standard deviation in the emission-line optical space, and also in the space of other physical parameters, allowing the development of astrophysical models that might be able to predict the physical conditions  responsible by the loci occupied by each class. 

\textit{Summa summarum}, this work takes a step forward in the systematic use of machine learning in astronomy. It provides a quantitative and robust recipe for unsupervised astronomical classification and how to combine its output with previous domain knowledge, hence conveying physically interpretable results.  Our approach stands out as a valuable tool  for future investigations, thanks to its potential to unveil non-trivial relationships in data which may be overlooked by standard procedures. Thus, we strongly advocate for the use of such techniques, especially due to their ability  to deal with high-dimensional datasets. 


 \section*{Acknowledgements}
 This work is a product of the $\rm 3^{rd}$ COIN Residence Program (CRP\#3). We thank Zsolt Frei for encouraging the accomplishment of this event. CRP\#3 was held in Budapest, Hungary on August 2016 and supported by E\"otv\"os University. \\ We specially thank E. E. O. Ishida for the  organization of the  CRP\#3 and the fruitful comments  and revision of the manuscript.   We thank K.~Parmar for the preliminary external cluster assessment script provided, and C.-A.Lin for the kind help provided during CRP\#3. The authors would also like to thank Prof. Pl\"{u}ssmaci for his useful feedback during the entire CRP\#3.
RSS thanks \emph{Fundação de Amparo à Pesquisa do Estado de São Paulo} (FAPESP) process no 2016/13470-3, and  2012/00800-4 for financial support. MLLD acknowledges \emph{Coordenação de Aperfeiçoamento de Pessoal de Nível Superior} (CAPES) and \emph{Instituto de Astronomia, Geofísica e Ciências Atmosféricas da Universidade de São Paulo} (IAG-USP) for the financial support. MLLD specially acknowledges S.~Rossi and \emph{Programa de Excelência Acadêmica} (PROEX) for the grant provided for attending CRP\#3 and, therefore, accomplishing this work. MLLD also thanks R.~Cid~Fernandes for providing the treated dataset, R.~Davies for the help provided and insights in the beginning of this work, and P.~Coelho for the fruitful discussions and support.
MVCD thanks his scholarship from FAPESP (processes 2014/18632-6 and 2016/05254-9). AKM thanks the Portuguese agency \emph{Funda\c c\~ao para Ci\^encia e Tecnologia} -- \emph{FCT}, for financial support (SFRH/BPD/74697/2010). RB was supported through the New National Excellence Program of the Ministry of Human Capacities, Hungary.
FG would like to acknowledge the generous support by the Radboud Excellence Initiative.

The IAA Cosmostatistics Initiative (COIN) is a non-profit organization whose aim is to nourish the synergy between astrophysics, cosmology, statistics, and machine learning communities.
 This work benefited from the following collaborative platforms: \texttt{Overleaf}\footnote{\url{https://www.overleaf.com}}, \texttt{Github}\footnote{\url{https://github.com}}, and \texttt{Slack}\footnote{\url{https://slack.com/}}.

In memoriam of Joseph M. Hilbe (30th December, 1944 - 12th March, 2017).

\appendix

\section{Internal Cluster Validation Methods}
\label{ap:diag}
\subsubsection*{Bayesian Information Criterion}

From a Bayesian viewpoint, model selection of a mixture model can be  estimated by the integrated likelihood  of the  model with $K$ components.  BIC  can be used as a  technique that penalizes the likelihood in model selection \citep{Schwarz1978,Liddle2007}.  The higher the value of BIC, the better the result. The BIC for a mixture model  log-likelihood is given  by 
\begin{equation}
BIC(K) = \log p(x|K,\hat{\theta}_k) - \frac{\nu_{K}}{2}\log{n}
\end{equation}
where $\hat{\theta}_k$ is the maximum likelihood estimate of $\theta_k$, and $\nu_{K}$ is the number of free parameters for a model with $K$ components.

\subsubsection*{Integrated Complete Likelihood}  

There is one  particular drawback when using BIC to find the best number of clusters: the method works 
appropriately when each mixture component corresponds to a separate cluster, but this is not always the case. In particular, a cluster may be both cohesive and well distanced from other clusters, without its distribution being Gaussian. Such cluster is best represented with two or more mixture components, rather
than a single Gaussian. Hence, the intrinsic number of data clusters may be different from the number of components in the  Gaussian mixture model.  \citet{Biernacki:2000} suggested an alternative to overcome this limitation by directly  estimating the number of clusters, as opposed to  the  number of mixture components; he proposed using the integrated complete likelihood (ICL),  which  can be roughly understood as BIC penalized by mean entropy \citep{Baudry2010}. As a rule of thumb, the number of clusters estimated by ICL is smaller than the number estimated by BIC, due to the additional entropy term.  We shall use both indices to constrain lower and upper  limits of our  solution.  
\subsubsection*{Entropy} 

A complementary visualization technique to validate  the number of clusters based on BIC and ICL is to use the \textit{elbow rule}: a graphical display of entropy variation against the number of clusters. The decrease of entropy at each step serves as a guideline to optimize the number of clusters \citep{Baudry2010}. 

\subsubsection*{Silhouette}    

The silhouette  approach measures the degree of similarity (dissimilarity) of objects within and between clusters \citep{rousseeuw1987}.  It quantifies the common sense that a good  clustering algorithm is able to partition the data such that the average distance between objects in the same cluster (i.e., the average intra-distance) is significantly lower
than the distance between objects in different clusters (i.e., the average inter-distance).  The technique assigns a value, known as the silhouette width, 
$s(i)$,  to a given cluster solution, which is defined as follows: 
\begin{equation}
s(i) = \frac{b(i)-a(i)}{\max{a(i),b(i)}},
\end{equation}
where $a(i)$ is the average distance between the $i^{th}$ object and all other objects in a given cluster; $b(i)$ is the minimum average distance between the objects in a given cluster and objects in other clusters.  Higher silhouette values indicate high-quality clustering solutions.  

\section{External Cluster Validation algorithm}
\label{app:cluster_val}

The methodology computes a distance matrix $M$ where rows correspond to classes and columns correspond to clusters. Each entry $M_{ij}$ captures the (probabilistic) distance between the two data groups. Each
class and cluster is modelled  as a multivariate Gaussian distribution $f(\textbf{x}) \sim N(\mu,\Sigma)$. 
From now one,  we refer to $f_i(\textbf{x})$ as the Gaussian model for a particular class $C_i$, and 
$f_j(\textbf{x})$ as the corresponding Gaussian model for a cluster $K_j$. We now describe the nature of 
the metric $\Psi(f_i,f_j)$ used to capture the distance between the two Gaussian models.  

A straightforward approach to measure the degree of separation $\Psi(f_i,f_j)$ between class $C_i$
and cluster $K_j$ is to use the concept of relative entropy (or Kullback–Leibler distance) of two density functions \cite{Cover2006}. 
The relative entropy  is the expectation 
of the logarithm of a likelihood ratio\footnote{The original definition contains $\log_2$, instead of
$\ln$; we prefer the latter because it simplifies when the functions are Gaussians; we switch then from a measurement in bits to one in nats.}.  

\begin{equation}
\Psi(f_i,f_j) = D(f_i || f_j) = \int_{\textbf{x}} f_i(\textbf{x}) \ln \frac{f_i(\textbf{x})}{f_j(\textbf{x})} d\textbf{x}.
\end{equation}
This measure can be interpreted as the error generated by assuming that $f_i(\textbf{x})$ can be used to 
represent $f_j(\textbf{x})$ (or alternatively, the additional amount of information required to describe $f_i(\textbf{x})$ given $f_j(\textbf{x}$). The higher the distance, the higher the dissimilarity between the two distributions. 

\begin{figure} 
\includegraphics[trim={0 0.25cm 0 2.25cm  },clip,width=\linewidth]
{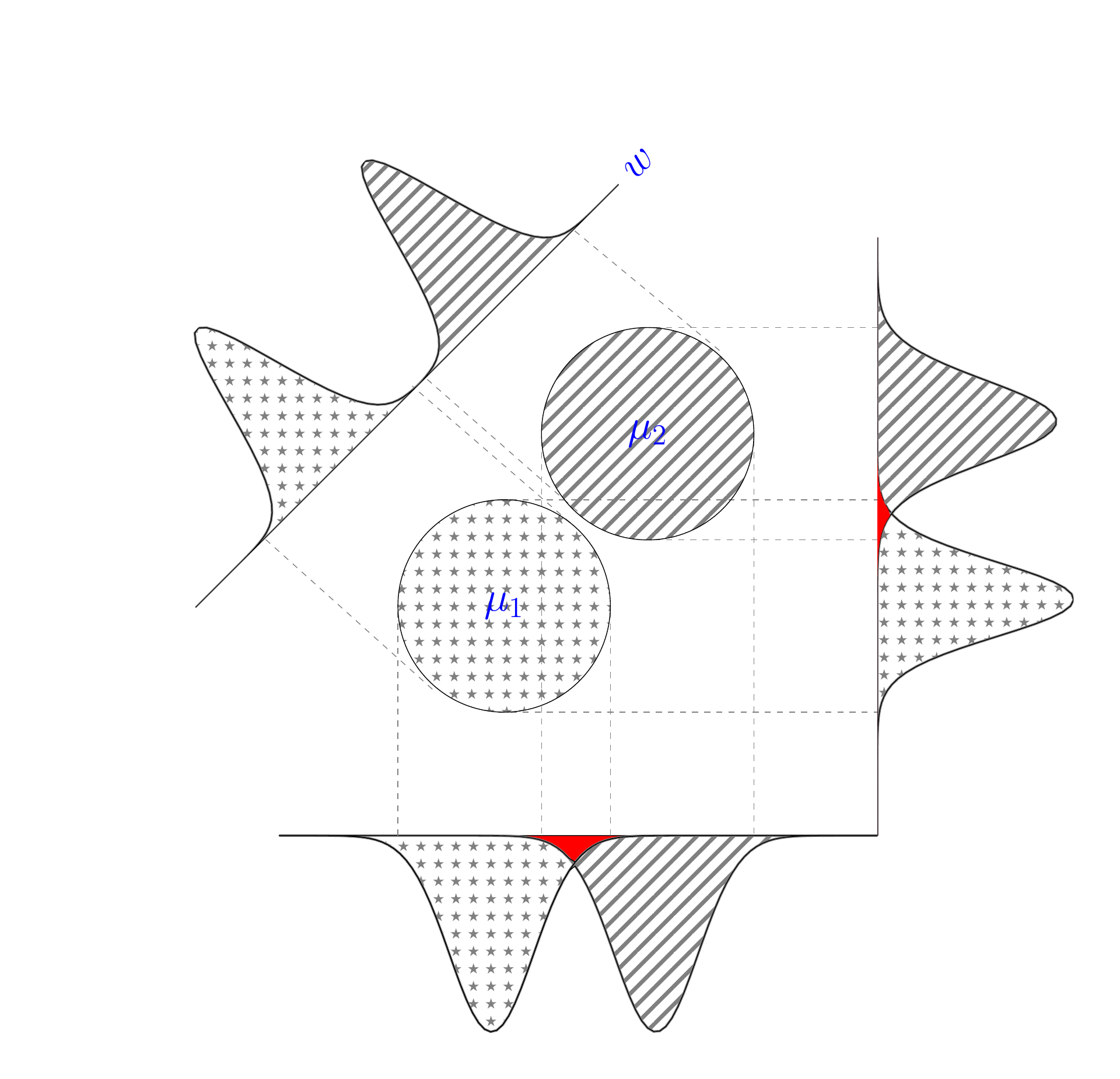}
\caption{Illustrative figure representing the linear discriminant analysis method. Weight vector $w$ which lies orthogonal to the hyperplane that maximizes the separation between the objects in cluster $K_j$ and  
class $C_i$ is used as the dimension over which galaxies are projected.}
\label{fig:fishersw}
\end{figure}

The metric defined above can be approximated using numerical methods, but the computational
cost can become very expensive; integrating over high-dimensional spaces soon turns intractable even for moderately low number of attributes. To address this problem,  one final step is necessary. 
The data is projected into a single dimension $w$, in order to  compute the distance function $\Psi$ along that 
dimension alone.  
In particular, the proposed solution consists of projecting data objects over a single dimension
that is orthogonal to Fisher's linear discriminant \citep{Duda2001,fisher1936}\footnote{The use of Fisher’s LDA is  appropriate here because both LDA and GMMs assume multivariate normality in the components}. The general idea
is to find a hyperplane that discriminates data objects in cluster $K_j$ from data objects
in class $C_i$. The weight vector $w$ that lies orthogonal to the hyperplane will be used as
the dimension upon which the data objects will be projected. The rationale behind this
method is that among all possible dimensions over which that data can be projected,
classical linear discriminant analysis identifies the vector $w$ with an orientation that
results in a maximum (linear) separation between data objects in $K_j$ and $C_i$; the distribution
of data objects over $w$ provide a better indication of the true overlap between $K_j$
and $C_i$ in multiple dimensions, compared to the resulting distributions obtained by projecting
data objects over the attribute axes. Figure~\ref{fig:fishersw} shows our methodology. Weight vector
$w$, which lies orthogonal to the hyperplane that maximizes the separation between
the objects in cluster $K_j$ and class $C_i$ is used as the dimension over which data objects
are projected.

To add more detail, Fisher's linear discriminant finds the vector $w$ that maximizes the following
criterion function: $J(w) = \frac{w^t S_B w}{w^t S_W w}$. $S_B$ is the 
between-class scatter matrix, defined as the outer product of two vectors: 
$S_B (\mu_j - \mu_i)^t (\mu_j - \mu_i)$, where $\mu_j$ and $\mu_i$ are the mean vectors of 
$f_j(\textbf{x})$ and $f_i(\mathrm{x})$ respectively.
$S_W$ is the within-class scatter matrix, defined as the scatter matrix over the two distributions: 
$S_W = \sum (\mathrm{x} - \mu_j)^t(\mathrm{x} - \mu_j)^t + \sum (\mathrm{x} - \mu_i)^t(\mathrm{x} - \mu_i)$. It can be shown that a solution maximizing $J(w)$ 
is in fact independent of $S_B$: $w = S^{-1}_W (\mu_j - \mu_i)$.
Geometrically the goal is to find a vector $w$ so that the difference of
the projected means over $w$ is large compared to the standard deviations around each
mean (Figure~\ref{fig:fishersw}).

\end{document}